\documentclass[onecolumn, draftcls]{IEEEtran}
\usepackage{amsmath,amssymb}
\usepackage{graphicx}
\usepackage{mathrsfs}
\usepackage{tikz}
\usepackage{booktabs}
\usepackage{enumerate}
\usepackage{psfrag}	
\usepackage{array}
\usepackage{multirow,hhline}
\usepackage{exscale}
\usepackage{color}
\usepackage{colortbl}
\usepackage{epsfig,subfigure}
\usepackage{cite}
\usepackage{amsthm}
\usepackage{hyperref}
\usepackage{dsfont}
\newcommand{\alert}[1]{\textcolor{black}{#1}}
\renewcommand{\vec}[1]{\ensuremath{\mathbf{#1}}}
\newtheorem{theorem}{Theorem}
\newtheorem{lemma}{Lemma}
\newtheorem{remark}{Remark}
\newtheorem{corollary}{Corollary}

\newtheorem{definition}{Definition}
\newtheorem{claim}{Claim}

\graphicspath{{./fig/}}

\begin{document}


\title{{Approximate Sum-Capacity of the Y-channel}}
\author{Anas~Chaaban, Aydin~Sezgin, and A.~Salman~Avestimehr
\thanks{A.~Chaaban and A.~Sezgin are with the Chair of Digital Communication Systems, Ruhr-Universit\"at Bochum (RUB), Universit\"atsstrasse 150, 44780 Bochum, Germany. Email: anas.chaaban@rub.de, aydin.sezgin@rub.de.}

\thanks{A.~S.~Avestimehr is with the School of Electrical and Computer Engineering Cornell University, 325 Frank H.~T.~Rhodes Hall, Ithaca, NY 14853, USA. Email: avestimehr@ece.cornell.edu.}

\thanks{The work of A. Chaaban and A. Sezgin is supported by the German Research Foundation, Deutsche
Forschungsgemeinschaft (DFG), Germany, under grant SE 1697/3. The work of A. S. Avestimehr is partly supported by NSF grants CAREER-0953117 and CCF-1161720.} 

\thanks{Part of the paper has been presented in Asilomar 2011 \cite{ChaabanSezginAvestimehr_Asilomar}.}}

\maketitle


\begin{abstract}
A network where three users want to establish multiple unicasts between each other via a relay is considered. This network is called the Y-channel and resembles an elemental ingredient of future wireless networks. The sum-capacity of this network is studied. A characterization of the sum-capacity within an additive gap of 2 bits, and a multiplicative gap of 4, for all values  of channel gains and transmit powers is obtained. Contrary to similar setups where the cut-set bounds can be achieved within a constant gap, they can not be achieved in our case, where they are dominated by our new genie-aided bounds. \alert{Furthermore, it is shown that a time-sharing strategy, in which at each time two users exchange information using coding strategies of the bi-directional relay channel, achieves the upper bounds to within a constant gap. This result is further extended to the K-user case, where it is shown that the same scheme achieves the sum-capacity within $2\log(K-1)$ bits.}
\end{abstract}

\begin{IEEEkeywords}
Multi-way relaying, sum-capacity, functional decode-and-forward, constant gap.
\end{IEEEkeywords}

\section{Introduction}
\IEEEpubidadjcol
\IEEEPARstart{M}{ulti}-way relaying promises becoming an essential ingredient of future networks. This is due to the rate enhancement arising from the possibility of using more sophisticated schemes in multi-way relaying setups, namely, schemes exploiting ideas from network coding.

A multi-way channel is a scenario where users communicate with each other in both directions. The simplest multi-way communication model is the two-way channel introduced by Shannon in \cite{Shannon_TWC} where 2 nodes communicate with each other, and each has a message to deliver to the other node. In this sense, each node is a source and a destination at the same time. The capacity of this two-way channel is not known in general.

The two-way channel can be extended to a bi-directional relay (BRC) channel by including a relay, where the two nodes communicate with each other via the relay. This setup was introduced in \cite{RankovWittneben} where relaying protocols were analyzed. Later, a comparison of the achievable rate regions of various relaying protocols for the BRC was preformed in \cite{KimDevroyeMitranTarokh}, where the authors used amplify-and-forward, decode-and-forward, compress-and-forward, and a mixture thereof. In a later development, the capacity of the downlink (broadcast) phase of the BRC was characterized in \cite{OechteringBjelakovicSchnurrBoche}. Furthermore, the capacity region for the Gaussian setting with a symmetric downlink was characterized within half a bit in \cite{GunduzTuncelNayak}. For the BRC with arbitrary channel parameters, the capacity region was characterized within half a bit, and the sum-capacity within $\log(3/2)$ bits in \cite{NamChungLee_IT}. The capacity region of the BRC was also characterized within a constant gap in \cite{AvestimehrSezginTse} with the aid of the linear-shift deterministic approximation \cite{AvestimehrDiggaviTse_IT} of the setup. The papers \cite{WilsonNarayananPfisterSprintson,NarayananPravinSprintson} also studied the BRC and obtained approximate sum-capacity characterizations using nested-lattice codes \cite{NazerGastpar}. \alert{For the purpose of this paper, we would like to draw the following conclusion from the aforementioned papers: \textit{The BRC has 2 degrees of freedom (DoF), which can be achieved with either physical-layer network coding (using nested-lattice coded), or a quantize-and-forward strategy.}}

The results for the BRC were also extended to the larger network consisting of two pairs of nodes in addition to the relay \cite{SezginKhajehnejadAvestimehrHassibi,SezginBocheAvestimehr}. The approximate capacity of the two-pair bi-directional relay network was obtained in \cite{SezginAvestimehrKhajehnejadHassibi,AvestimehrKhajehnejadSezginHassibi}.

If more than two nodes want to communicate via a relay in a multi-directional manner, the multi-way relay channel (MRC) is obtained. The MRC was studied in \cite{GunduzYenerGoldsmithPoor_IT}, where upper and lower bounds for the capacity of the Gaussian setting were given. In this setting, G\"und\"uz {\it et al.} divided users into several clusters, where each user in a cluster has a single message intended to all other users in the same cluster (multicast). A similar multicast setup with all users belonging to the same cluster, and all channel gains, noise variances, and power constraints being equal, was considered in \cite{OngKellettJohnson_IT} where the sum-capacity was obtained for the $k>2$ user case. This provides the DoF of this multicast setup. For instance, for the 3-user multicast MRC, the DoF is 3/2 (less than that of the BRC).

\IEEEpubidadjcol

\subsection{Main Contribution}
In this paper, we consider a 3-user Gaussian MRC, with a notable difference from the aforementioned MRC's. Namely, 3 users communicate with each other simultaneously via a relay, where each user has 2 independent messages, each of which is intended to one of the other users (unicast). Thus each node wants to send 2 messages to the other nodes, and wants to decode 2 other messages. This setup generalizes the BRC to 3 users. 
It first appeared in a MIMO setting in \cite{LeeLim}, where a transmission scheme based on interference alignment was proposed, and its corresponding achievable DoF were calculated. It was referred to as the ``Y-channel". In \cite{LeeLimChun}, it was shown that if the relay has at least $\lceil{3M/2}\rceil$ antennas where $M$ is the number of antennas at the other nodes, then the cut-set bound is asymptotically achievable. That is, the cut-set bound characterizes the DoF of the setup in this case. \alert{However, Lee {\it et al.} left the DoF of the case where the relay has less than $\lceil{3M/2}\rceil$ antennas open, and hence the SISO case is also left open.}\footnote{We refer to the SISO Y-channel simply as the Y-channel for brevity.} \alert{Our main contribution in this paper is a constant gap characterization of the sum-capacity of the Y-channel. In the discussion of the results, we use the DoF as a comparison criterion for different bounds and networks, and as a stepping stone towards constant gap results.} Our contributions to this problem can be summarized as follows:

\subsubsection{Upper Bounds}
We consider the single antenna Gaussian Y-channel where all nodes are full-duplex. We distinguish between two cases: a non-restricted Y-channel, and a restricted Y-channel. In the non-restricted case, the transmit signals of the users can depend on the previously received symbols, while in the restricted case they can not. If the relay is equipped with multiple antennas, a DoF of 3 can be achieved~\cite{LeeLimChun} as the spatial dimensions offer additional signaling space. On the other hand, a relay equipped with a single antenna might represent a bottleneck in the system with an impact on the DoF, which has not been characterized so far. 
Here, we address this problem.
We derive new upper bounds for the sum-capacity of this channel using a genie-aided approach. While the cut-set bounds provide a DoF upper bound of 3, we provide a DoF upper bound of 2, which is smaller than that obtained from the cut-set bounds. It follows that the cut-set bounds do not characterize the DoF in our case, contrary to the MIMO case in \cite{LeeLimChun}, \alert{nor do they provide a constant gap characterization of the sum-capacity.}

\subsubsection{Lower Bounds}
In \cite{OngKellettJohnson_IT}, the so-called ``functional decode-and-forward" (FDF) scheme was used as an achievable scheme for the MRC. However, in \cite{OngKellettJohnson_IT}, the multicast case was considered. We modify the FDF scheme accordingly to obtain a lower bound for the sum-capacity using network coding and nested-lattice codes. The main idea of FDF is to allow the relay to decode a function of the transmitted codewords, and then forward it to the users in a way that each user is enabled to extract his desired messages. The relay can also decode the codewords sent by the users ``completely'' (instead of a function of the codewords) and then forward this information to the users. This scheme is called ``complete decode-and-forward'' (CDF) and is used to obtain another lower bound on the sum-capacity.

\subsubsection{Assessment}
\alert{A DoF of 2 can be achieved in the Y-channel by operating it as a BRC. {\it As a consequence of our new upper bound, we conclude that the DoF of the Y-channel is in fact equal to 2, and that the DoF-cut-set bound is not achievable.}} 
\alert{This conclusion is further extended to the $K$-user case, where each of the $K$ users has $K-1$ independent messages, each of which is intended to a different user. This channel, the $K$-user star channel, also has 2 DoF, same as the Y-channel and the BRC.} 

\alert{The intuition behind this result is as follows. Since 2 DoF are achievable in the BRC, then 2 DoF are also achievable in the Y-channel. Now increasing the number of users in a network can not decrease the DoF (one can always switch some users off), but it can increase the DoF (the DoF of the interference channel increase with the number of users \cite{CadambeJafar_KUserIC}). In our setup however, the relay is a bottleneck for the DoF. The relay, having 1 antenna, can only obtain 1 (noisy) equation, and hence, can resolve only 1 interference free symbol. In classical relay networks, this would account for 1 DoF. In multi-way relaying setups however, this accounts for 2 DoF due to the possibility of network coding. This motivates finding upper bounds which reflect this 2 DoF behavior, and hence showing that the Y-channel and the BRC have the same DoF.}

By comparing the upper and lower bounds for the non-restricted Y-channel, we bound the gap between them. This gap is shown to be less than 2 bits for all values of channel gains. We also bound the multiplicative gap between the bounds, i.e., the ratio of the upper bound to the lower bound, by a factor of 4. For the symmetric Y-channel where all channel gains are equal and for the restricted Y-channel with arbitrary channel gains, the bounds can be further tightened and we characterize the sum-capacity within 1 bit. For the K-user case, the sum-capacity is characterized within $2\log(K-1)$ bits. Interestingly, the same scheme which achieves the approximate sum-capacity of the BRC achieves the approximate sum-capacity of the K-user case too.

The rest of the paper is organized as follows. The system model is described in Section \ref{Model}. The non-restricted Y-channel is considered first, and upper bounds for its sum-capacity are given in Section \ref{UpperBounds} and lower bounds in Section \ref{LowerBounds}. The gap between upper and lower bounds for the non-restricted Y-channel is calculated in Section \ref{GapCalculation:General}, and for the restricted Y-channel in Section \ref{Section:RYC}. Extensions of these results are considered in Section \ref{Sec:Extensions} and we conclude the paper with Section \ref{Conclusion}. 

Throughout the paper, we use the following notation. $x^n$ denotes a sequence of $n$ symbols $(x_1,\cdots,x_n)$. $\mathcal{N}(\mu,\sigma^2)$ is used to denote a Gaussian distribution with mean $\mu$ and variance $\sigma^2$. We use $[i:j]$ to denote $\{i,i+1,\dots,j\}$, $C(x)=\frac{1}{2}\log(1+x)$ and $C^+(x)=\max\{0,C(x)\}$. Symbols with an overline $\overline{x}$ or an underline  $\underline{x}$ denote an upper bound or a lower bound, respectively.

\section{System Model}
\label{Model}

\begin{figure}
\centering
\includegraphics[width=.5\columnwidth]{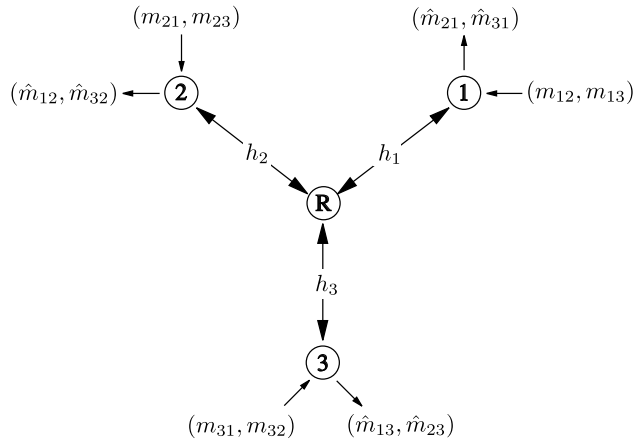}
\caption{The Y-channel: User 1 wants to send two messages, $m_{12}$ to user 2, and $m_{13}$ to user 3. User 1 also wants to decode two messages, $m_{21}$ from user 2, and $m_{31}$ from user 3. Similarly at users 2 and 3.}
\label{Fig:Model}
\end{figure}

The Y-channel models a setup where 3 users want to communicate with each other in a multi-directional manner via a relay as shown in Figure \ref{Fig:Model}. Each user has an independent message to each other user. Consequently, each user wants to send two messages via the relay to the two remaining users, and wants to decode two messages. We assume that all nodes are full-duplex, and that there is an AWGN channel between each node and the relay and vice versa. That is, there is a forward and a backward AWGN channel between each node and the relay, where the noise is of zero-mean and unit-variance.

User $j$ has messages 
\begin{align*}
m_{jk}&\in\mathcal{M}_{jk}\triangleq\{1,\dots,2^{nR_{jk}}\}, 
\end{align*}
and
\begin{align*}
m_{jl}&\in\mathcal{M}_{jl}\triangleq\{1,\dots,2^{nR_{jl}}\},
\end{align*}
to be delivered to users $k$ and $l$, respectively, where $R_{jk},R_{jl}\in\mathbb{R}_+$, for all distinct $j,k,l\in\{1,2,3\}$. Thus, there are six mutually independent messages in total. 

\begin{remark}
Notice that due to this independence assumption, the results of this paper do not cover the setup in \cite{GunduzYenerGoldsmithPoor_IT, OngKellettJohnson_IT}. The considered setup in those papers corresponds to setting $m_{jk}=m_{jl}$ in our setup, which is not possible due to the independence of the messages.
\end{remark}

The messages of user $j$, $m_{jk}$ and $m_{jl}$, are encoded into a sequence $x_j^n$ using an encoding function, where $x_{ji}$ ($i=1,\dots,n$), the $i$th symbol of $x_j^n$, is a realization of a real random variable $X_{ji}$ satisfying the following power constraint 
\begin{align*}
\frac{1}{n}\sum_{i=1}^n\mathbb{E}[X_{ji}^2]\leq P.
\end{align*}
In general, the codeword $x_j^n$ is generated using an encoding function which maps the messages $m_{jk}$ and $m_{jl}$, and the previously received symbols at node $j$ till time instant $i$, $y_j^{i-1}$, to $x_{ji}$, i.e., 
\begin{align}
\label{SourceEncoder}
x_{ji}=f_{ji}(m_{jk},m_{jl},y_j^{i-1}).
\end{align}
This is the most general case and is called {\it non-restricted encoding}. {\it Restricted encoding} in contrast only uses the messages $m_{jk}$ and $m_{jl}$ to construct $x_{ji}$, i.e.,
\begin{align}
\label{SourceEncoder_R}
x_{j}^n=f_j(m_{jk},m_{jl}).
\end{align}
{These two cases have been pointed out by Shannon in his work on the two-way channel \cite{Shannon_TWC}.} In the Y-channel with non-restricted encoding, which we call a \textit{non-restricted Y-channel}, the transmit signals of different users can be dependent. This is not the case with restricted encoding in what we call a \textit{restricted Y-channel}.

The received signal at the relay at time instant $i$ is given by
\begin{align*}
y_{ri}=h_1x_{1i}+h_2x_{2i}+h_3x_{3i}+z_{ri},
\end{align*}
where $z_{ri}$ is a realization of an independent and identically distributed (i.i.d.) Gaussian noise $Z_r\sim\mathcal{N}(0,1)$ and $h_1,h_2,h_3\in\mathbb{R}$ are the constant channel coefficients from the users to the relay. We assume without loss of generality that 
\begin{align}
\label{Ordering}
h_1^2\geq h_2^2\geq h_3^2,
\end{align}
i.e., user 1 is the strongest user, and user 3 is the weakest. The relay sends a sequence $x_r^n$ of random variables $X_{ri}$ that satisfy the power constraint
\begin{align*}
\frac{1}{n}\sum_{i=1}^n\mathbb{E}[X_{ri}^2]\leq P,
\end{align*}
which depends on the past received symbols at the relay till time instant $i$, i.e.,
\begin{align}
\label{RelayEncoder}
x_{ri}=f_{ri}(y_r^{i-1}).
\end{align}
Then, the received signal at user $j$ and time instant $i$ can be written as
\begin{align}
\label{ReceivedSignal}
y_{ji}=h_jx_{ri}+z_{ji},
\end{align}
where $z_{ji}$ is a realization of an i.i.d. Gaussian noise $Z_j\sim\mathcal{N}(0,1)$. The channel is assumed to be reciprocal\footnote{the non-reciprocal case is discussed briefly in Section \ref{Sec:NonReciprocal}}, i.e., the channel gain from user $j$ to the relay is the same as that from the relay to user $j$. Each node $j$ uses a decoding function $g_j$ to decode $m_{kj}$ and $m_{lj}$ from the information available at that node, i.e.,
\begin{align*}
(\hat{m}_{kj},\hat{m}_{lj})=g_j(y_j^n,m_{jk},m_{jl}).
\end{align*}
For simplicity of exposition, we need the following definition.

\begin{definition}
We denote the vector of all rates by $\vec{R}$
and that of all messages by $\vec{m}$,
\begin{align*}
\vec{R}&=(R_{12},R_{13},R_{21},R_{23},R_{31},R_{32}),\\
\vec{m}&=(m_{12},m_{13},m_{21},m_{23},m_{31},m_{32}).
\end{align*}
We also define $R_\Sigma(\vec{R})$ to be the sum of the components of $\vec{R}$,
\begin{align*}
R_\Sigma(\vec{R})=\sum_{j=1}^3\sum_{\substack{k=1\\ k\neq j}}^3R_{jk}.
\end{align*}
\end{definition}

The message sets $\mathcal{M}_{jk}$, encoding functions $f_{ji}$ (or $f_j$ for the restricted Y-channel), $f_{ri}$, and decoding functions $g_j$ define a code $(\vec{R},n)$ for the Y-channel. A decoding error occurs if $(\hat{m}_{kj},\hat{m}_{lj})\neq({m}_{kj},{m}_{lj})$, for some distinct $j,k,l\in\{1,2,3\}$. A rate tuple $\vec{R}\in\mathbb{R}_+^6$ is said to be achievable if there exists a sequence of $(\vec{R},n)$ codes with an average error probability that approaches zero as $n$ increases. The set of all achievable rate tuples is the capacity region $\mathcal{C}$ of the Y-channel, and an achievable sum-rate is defined as $R_\Sigma(\vec{R})$ where $\vec{R}\in\mathcal{C}$. The sum-capacity is the maximum achievable sum-rate given by
\begin{align*}
C=\max_{\vec{R}\in\mathcal{C}}R_\Sigma(\vec{R}).
\end{align*}
For simplicity, we denote $R_\Sigma(\vec{R})$ by $R_\Sigma$, and we denote the ratio of available transmit power $P$ to noise power by $\mathsf{SNR}$. The DoF is defined as $$d=\lim_{\substack{\mathsf{SNR}\to\infty}}\frac{C}{\frac{1}{2}\log(\mathsf{SNR})}.$$
\alert{While the main results of the paper involve capacity characterization within a constant gap, the DoF will be used as a measure for comparing different bounds in order to judge their tightness, and as a stepping stone towards approximate capacity characterization.}

In the following sections, we will study the sum-capacity of the Y-channel, by deriving upper and lower bounds. Then we bound the gap between the upper and lower bounds. We consider both the non-restricted Y-channel where the encoding functions are as given in (\ref{SourceEncoder}) whose sum-capacity will be denoted {$C_N$}, and the restricted Y-channel where the encoding functions are as given in (\ref{SourceEncoder_R}) whose sum-capacity will be denoted {$C_R$}. Clearly, $C_R\leq C_N$.

\section{The Non-restricted Y-channel: Upper Bounds}
\label{UpperBounds}
We start by considering the non-restricted Y-channel, and give sum-capacity upper bounds for this case. Note that an upper bound for the non-restricted Y-channel is also an upper bound for the restricted one. 

\subsection{Cut-set Bounds}
\begin{figure}[t]
\centering
\includegraphics[width=.5\columnwidth]{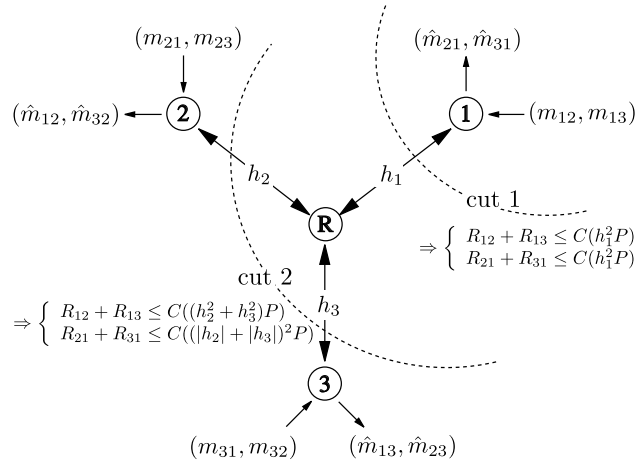}
\caption{{Cut-set bounds for the Y-channel. Cut 1 splits the set $\mathcal{S}=\{\text{user 1},\text{user 2},\text{user 3},\text{relay}\}$ into $\mathcal{T}_1=\{\text{user 1}\}$ and $\mathcal{T}_1^c$. This can be used to obtain a bound on $R_{12}+R_{13}$ if we consider information flow from $\mathcal{T}_1$ to $\mathcal{T}_1^c$, and on $R_{21}+R_{31}$ if we consider information flow from $\mathcal{T}_1^c$ to $\mathcal{T}_1$. Similarly, using cut 2 we can obtain one more bound on both $R_{12}+R_{13}$ and $R_{21}+R_{31}$.}}
\label{Cut}
\end{figure}

One can obtain upper bounds for the Y-channel by using the cut-set bounds {\cite{CoverThomas}}. If we label the set of nodes in the Y-channel by $\mathcal{S}\triangleq\{\text{user 1},\text{user 2},\text{user 3},\text{relay}\}$, then the cut-set bounds provide upper bounds on the rate of information flow from a set $\mathcal{T}\subset\mathcal{S}$ (such that $\text{user j}\in\mathcal{T}$ for some $j\in\{1,2,3\}$) to its complement $\mathcal{T}^c$ in $\mathcal{S}$ (see Figure \ref{Cut}). The cut-set bounds for the MIMO Y-channel were derived in \cite{LeeLimChun}. The cut-set bounds for the single antenna Y-channel are provided in the following theorem.

\begin{theorem}
\label{CutSetBounds}
\alert{If a rate tuple $\vec{R}$ is achievable, then there exists some joint probability distribution $p(x_1,x_2,x_3,x_r)$ satisfying the power constraints such that}
\begin{align}
\label{CS1}
&\hspace{-0.3cm}R_{jk}+R_{jl}\leq\min\left\{I(X_j;Y_r|X_k,X_l,X_r),\right.\nonumber\\
&\hspace{4.5cm}\left.I(X_r;Y_k,Y_l|X_k,X_l)\right\}\\
\label{CS2}
&\hspace{-0.3cm}R_{jl}+R_{kl}\leq\min\left\{I(X_j,X_k;Y_r|X_l,X_r),I(X_r;Y_l|X_l)\right\}
\end{align}
for all distinct $j,k,l\in\{1,2,3\}$.
\end{theorem}
\begin{IEEEproof}
The bounds are obtained using the standard cut-set bound approach \cite{CoverThomas}. The first bound (\ref{CS1}) in Theorem \ref{CutSetBounds} is obtained from the cut-sets $\mathcal{T}=\{\text{user j}\}$ and $\mathcal{T}=\{\text{user j},\text{relay}\}$, respectively. The last bound (\ref{CS2}) in Theorem \ref{CutSetBounds} is obtained from the complementary cut-sets. The rates to/from user 1 for instance can be bounded using the cut-sets shown in Figure \ref{Cut} labeled as cut 1 and cut 2. Let us start with cut 1. Using this cut, we can bound the rate of information flow from the set $\mathcal{T}_1=\{\text{user 1}\}$ to its complement $\mathcal{T}_1^c=\{\text{user 2}, \text{user 3}, \text{relay}\}$ by
\begin{align*}
R_{12}+R_{13}\leq I(X_1;Y_r|X_2,X_3,X_r),
\end{align*}
{and in the other direction, i.e., from $\mathcal{T}_1^c$ to $\mathcal{T}_1$, we get}
\begin{align*}
R_{21}+R_{31}\leq I(X_r;Y_1|X_1).
\end{align*}
{On the other hand, using cut 2, we can bound the rate of information flow from the set $\mathcal{T}_2=\{\text{user 1}, \text{relay}\}$ to its complement $\mathcal{T}_2^c=\{\text{user 2}, \text{user 3}\}$ by}
\begin{align*}
R_{12}+R_{13}\leq I(X_r;Y_2,Y_3|X_2,X_3),
\end{align*}
{and in the other direction, i.e., from $\mathcal{T}_2^c$ to $\mathcal{T}_2$, we get}
\begin{align*}
R_{21}+R_{31}\leq I(X_2,X_3;Y_r|X_1,X_r).
\end{align*}
\end{IEEEproof}
The following bounds are obtained as a corollary from Theorem \ref{CutSetBounds}.

\begin{corollary}
\label{CSG}
The achievable rates in the Y-channel satisfy
\begin{align}
\label{CSG1}
R_{jk}+R_{jl}&\leq C\left(\min\left\{h_j^2,h_k^2+h_l^2\right\}P\right)\\
\label{CSG2}
R_{jl}+R_{kl}&\leq C\left(\min\left\{(|h_j|+|h_k|)^2,h_l^2\right\}P\right),
\end{align}
for all distinct $j,k,l\in\{1,2,3\}$.
\end{corollary}
\begin{IEEEproof}
The bounds are obtained by maximizing the cut-set bounds individually using the Gaussian distribution for the inputs $(X_1,X_2,X_3,X_r)$ as shown in Appendix \ref{CSGProof}.
\end{IEEEproof}

\alert{In order to derive the sum-capacity upper bound that follows from Corollary \ref{CSG}, one would need to maximize $R_\Sigma$ under the constraints in the corollary. However, it can clearly be seen that the bounds in Corollary \ref{CSG} translate to 1 DoF each. Thus, by maximizing the sum DoF under the DoF constraints resulting from the corollary\footnote{\alert{This is a linear program which can be solved using the simplex method~\cite{MatousekGartner}.}}, we get $$d\leq3,$$ an upper bound whose looseness is to be proved in the next subsection. We conclude this subsection with the following sum-capacity upper bound obtained from Corollary \ref{CSG}.}

\begin{corollary}
\label{Corollary:CutSetBounds}
The sum-capacity of the Y-channel {with $h_1^2\geq h_2^2\geq h_3^2$} is upper bounded by $\overline{C}_{CS}$ where
\begin{align*}
\overline{C}_{CS}\triangleq C(\min\{h_1^2,h_2^2+h_3^2\}P)+C(h_2^2P)+C(h_3^2P),
\end{align*}
where the subscript $()_{CS}$ is used to denote the bound obtained from the cut-set bounds.
\end{corollary}
\begin{IEEEproof}
From Corollary \ref{CSG}, by evaluating \eqref{CSG1} and using \eqref{Ordering} we have
\begin{align}
\label{CSBnd1}
R_{12}+R_{13}&\leq C(\min\{h_1^2,h_2^2+h_3^2\}P),\\
\label{CSBnd2}
R_{21}+R_{23}&\leq C(\min\{h_2^2,h_1^2+h_3^2\}P)=C(h_2^2P),\\
\label{CSBnd3}
R_{31}+R_{32}&\leq C(\min\{h_3^2,h_1^2+h_2^2\}P)=C(h_3^2P).
\end{align}
Adding up these bounds, we obtain
\begin{align*}
R_\Sigma&\leq C(\min\{h_1^2,h_2^2+h_3^2\}P)+C(h_2^2P)+C(h_3^2P),
\end{align*}
and the statement of the corollary follows.
\end{IEEEproof}

Although $\overline{C}_{CS}$ might not be the tightest sum-capacity bound obtained from Corollary \ref{CSG}, it suffices for the purpose of the paper\footnote{We do not claim that the bound $\overline{C}_{CS}$ is the tightest bound that can be obtained from Corollary \ref{CSG}. However, it is not possible to use this corollary to obtain a sum-capacity upper bound with a lower pre-log than $\overline{C}_{CS}$ since the DoF-cut-set bound is $d\leq3$. This means that the tightest sum-capacity bound that can be obtained from Corollary \ref{CSG} provides only a minor improvement compared to $\overline{C}_{CS}$.}. It additionally reflects the DoF-cut-set bound $d\leq3$.

Having this upper bound, one could ask whether this upper bound is asymptotically achievable. That is, is it possible to achieve 3 DoF in the Y-channel? The answer of this question is given in the following subsection.

\subsection{Genie-aided Bounds}
Recall that in \cite{GunduzYenerGoldsmithPoor_IT} and \cite{OngKellettJohnson_IT}, the multicast setting of the MRC was studied, where each node has a single message intended to all other nodes. In that case, it was shown that the cut-set bounds are sufficient to obtain an asymptotic characterization of the sum-capacity. Furthermore, in \cite{LeeLimChun} where the MIMO Y-channel was considered, it was shown that if the relay has more than $\lceil{3M/2}\rceil$ antennas, where $M$ is the number of antennas at the other nodes, then the sum-capacity bound obtained from the cut-set bound is asymptotically achievable. Interestingly however, in our setting this is not the case\footnote{Note that the condition on the number of antennas in \cite{LeeLimChun} is not satisfied in our case.}. 

Corollary \ref{CSG} bounds the sum of 2 rates (2 components of $\vec{R}$) by $\frac{1}{2}\log(P)+o(\log(P))$, providing a DoF upper bound of 3. It turns out that this is not the true DoF of the Y-channel. In the following, we develop a tighter upper bounds, and show that the Y-channel has a DoF of 2. The key is to find a method to bound the sum of 3 rates (3 components of $\vec{R}$) by $\frac{1}{2}\log(P)+o(\log(P))$. If this is possible, then a DoF upper bound of 2 is obtained, and hence, a tighter bound than the cut-set bounds.
This tighter bound is given in Theorem \ref{SRUBG}, but before we state this theorem, we need the following lemmas.

\begin{lemma}
\label{FromRelay}
The achievable rates in the Y-channel 
must satisfy
\begin{align*}
R_{kj}+R_{lj}+R_{kl}&\leq C((h_j^2+h_l^2)P)
\end{align*}
for all distinct $j,k,l\in\{1,2,3\}$.
\end{lemma}
\begin{IEEEproof}
This bound is obtained by using a genie-aided approach to bound the sum of three rates. Details are given in Appendix~\ref{ProofFromRelay}.
\end{IEEEproof}

The idea here is to generate two upper bounds on $R_{21}+R_{31}$ and on $R_{23}$ for instance, which have have the following form 
\begin{align*}
R_{21}+R_{31}&\leq I(A;B),\\
R_{23}&\leq I(A;C|B),
\end{align*}
so that the two bounds can be combined using the chain rule of mutual information, such that the resulting mutual information expression $I(A;B,C)$ is equivalent to 1 DoF ($\frac{1}{2}\log(P)+~o(\log(P))$). This can be accomplished by carefully designing a genie-aided Y-channel and using it to obtain such bounds. See Appendix \ref{ProofFromRelay} for more details. Another similar upper bound is provided in the following lemma.

\begin{lemma}
\label{GUB}
The achievable rates in the Y-channel must satisfy
\begin{align*}
R_{kj}+R_{lj}+R_{kl}&\leq C((|h_k|+|h_l|)^2P)
\end{align*}
for all distinct $j,k,l\in\{1,2,3\}$.
\end{lemma}
\begin{IEEEproof}
The proof is given in Appendix \ref{GeneralProof}.
\end{IEEEproof}

As a result of Lemmas \ref{FromRelay} and \ref{GUB}, we obtain 
\begin{align}
&R_{kj}+R_{lj}+R_{kl}\nonumber\\
\label{3BoundG}
&\hspace{0.5cm}\leq \min\left\{C((h_j^2+h_l^2)P),C((|h_k|+|h_l|)^2P)\right\}.
\end{align}
Consequently, we obtain the following theorem.

\begin{theorem}
\label{SRUBG}
The sum-capacity of the Y-channel is upper bounded by $\overline{C}_{N}$, i.e.
\begin{align}
\label{CgBound}
C_N\leq\overline{C}_{N}&=C((h_2^2+h_3^2)P)\\
&\quad+C(\min\{h_1^2+h_3^2,(|h_2|+|h_3|)^2\}P)\nonumber.
\end{align}
\end{theorem}
\begin{IEEEproof}
By evaluating (\ref{3BoundG}) for $(j,k,l)=(2,1,3)$, and for $(j,k,l)=(1,3,2)$ and  adding the two obtained bounds, we arrive at the desired result.
\end{IEEEproof}

\begin{remark} 
Note that \eqref{3BoundG} can be evaluated for different tuples $(j,k,l)$ as well to get upper bounds. However, they are not more binding than \eqref{CgBound}.
\end{remark}

Now, recall our discussion before Lemma 1 on the DoF of the Y-channel. As we can see, the bound in Theorem \ref{SRUBG} provides a DoF upper bound of $$d\leq2,$$ clearly tighter than the DoF-cut-set bound. Therefore, the sum-capacity upper bound $\overline{C}_N$ becomes tighter than $\overline{C}_{CS}$ as $P$ increases, and \alert{hence, \textit{the cut-set bound is not asymptotically achievable}. Moreover, since the BRC has 2 DoF, achievable by network coding using nested-lattice codes, then 2 DoF are also achievable in the Y-channel by operating it as a BRC. As a result, the Y-channel has 2 DoF, same as the BRC.}

\alert{As can be seen, by extending the BRC to 3 users, maintaining full message exchange (multiple unicasts), the DoF do not change and are still achievable with network coding and lattice codes. Does this result hold also for an arbitrary number of users? This question is answered in the affirmative in Section \ref{K-user-Star}.}

Next, we provide various transmission schemes for the Y-channel where we use complete decode-and-forward and functional decode-and-forward.

\section{Transmission Schemes and Lower Bounds}
\label{LowerBounds}
\alert{Several transmission schemes can be used for the Y-channel, including decode-and-forward and compute-and-forward \cite{NazerGastpar}. Before proceeding, we would like to note that the schemes we provide apply for both the non-restricted and the restricted Y-channel, since all the schemes we provide use the restricted encoder \eqref{SourceEncoder_R} (and not \eqref{SourceEncoder}) for encoding the messages at all the three users.}

\subsection{Complete Decode-and-Forward}
\label{LowerBound1}
{Complete decode-and-forward (CDF) refers to the scheme where the relay `completely' decodes all the messages sent by the users and then forwards them. This is, for instance, in contrast with `partial' decode-and-forward where the relay only decodes some messages, or `functional' decode-and-forward where the relay decodes a function of the sent messages.}

{Before we proceed with explaining this scheme, we start by summarizing CDF for the BRC \cite{OngKellettJohnson_IT}. In this scheme, the relay decodes all the sent messages (two in this case), and then encodes all the decoded messages into one signal $x_r^n$ to be sent in the next transmission block. Each user decodes all the unknown messages (one message in the two user case) given his own message as side information. This is shown to achieve the following region \cite{Knopp,NamChungLee} (assuming that the channel gains are $h_1$ and $h_2$ with $h_2^2\leq h_1^2$)}
\begin{align}
\label{2WRCCDF}
\left\{(R_1,R_2)\in\mathbb{R}^2_+\left|\begin{array}{rl}
R_1,R_2&\leq C\left(h_2^2P\right)\\
R_1+R_2&\leq C((h_1^2+h_2^2)P)\end{array}\right.\right\}.
\end{align}
In what follows, we extend this scheme to the Y-channel starting with the uplink.

\subsubsection{Uplink}
In this scheme, user $j$ encodes his messages $m_{jk}$ and $m_{jl}$ into an i.i.d. sequence $x_j^n(m_{jk},m_{jl})$ where $X_j\sim\mathcal{N}(0,P)$. Then, all users transmit their signals to the relay simultaneously. The relay decodes all messages in a MAC fashion. That is, we treat the uplink channel as a MAC channel where the relay decodes the signals $x_1^n$, $x_2^n$, and $x_3^n$ successively or jointly (simultaneous decoding \cite{ElgamalKim}). Thus, the following rate region is achievable in the uplink:
\begin{align}
\label{MACRC1}
R_{12}+R_{13}&\leq C(h_1^2P)\\
\label{MACRC2}
R_{21}+R_{23}&\leq C(h_2^2P)\\
\label{MACRC3}
R_{31}+R_{32}&\leq C(h_3^2P)\\
\label{MACRC4}
R_{12}+R_{13}+R_{21}+R_{23}&\leq C((h_1^2+h_2^2)P)\\
\label{MACRC5}
R_{12}+R_{13}+R_{31}+R_{32}&\leq C((h_1^2+h_3^2)P)\\
\label{MACRC6}
R_{21}+R_{23}+R_{31}+R_{32}&\leq C((h_2^2+h_3^2)P)\\
\label{MACconstraint}
R_\Sigma&\leq C((h_1^2+h_2^2+h_3^2)P).
\end{align}

\subsubsection{Downlink: Variant 1}
A straight forward extension of the CDF scheme of the two-way relay channel is as follows. After the relay decodes $\vec{m}$ from its received signal, it uses a Gaussian codebook to encode $\vec{m}$ into an i.i.d. sequence $x_r^n(\vec{m})$ where $X_r\sim\mathcal{N}(0,P)$. The relay then sends $x_r^n(\vec{m})$. The problem now resembles a broadcast problem with side information at the receivers \cite{Tuncel,KramerShamai}. After receiving a noisy observation of $x_r^n(\vec{m})$, user 1 knowing $m_{12}$ and $m_{13}$ can decode all other messages as long as \cite{GunduzYenerGoldsmithPoor_IT}
\begin{align}
\label{DFC1}
R_{21}+R_{23}+R_{31}+R_{32}\leq C(h_1^2P).
\end{align}
Similarly at the other receivers, reliable decoding is guaranteed if the following rate constraints are fulfilled
\begin{align}
\label{DFC2}
R_{12}+R_{13}+R_{31}+R_{32}&\leq C(h_2^2P),\\
\label{DFC3}
R_{12}+R_{13}+R_{21}+R_{23}&\leq C(h_3^2P).
\end{align}
Collecting the rate constraints \eqref{MACRC1}-\eqref{DFC3}, and removing all redundant terms, we can see that a rate tuple $\vec{R}$ that satisfies the following rate constraints
\begin{align}
\label{CDFV1B1}
R_{31}+R_{32}&\leq C(h_3^2P)\\
\label{CDFV1B2}
R_{12}+R_{13}+R_{21}+R_{23}&\leq C(h_3^2P)\\
R_{12}+R_{13}+R_{31}+R_{32}&\leq C(h_2^2P)\\
R_{21}+R_{23}+R_{31}+R_{32}&\leq C(\min\{h_1^2,h_2^2+h_3^2\}P)\\
\label{CDFV1B5}
R_\Sigma&\leq C((h_1^2+h_2^2+h_3^2)P),
\end{align}
is achievable. In order to find the maximum achievable sum-rate for CDF, we solve
\begin{align}
\label{OptProb}
&\text{maximize}\quad \sum_{j=1}^3\sum_{\substack{k=1\\k\neq j}}R_{jk}\\
&\text{subject to}\quad \nonumber\\
&R_{jk}\geq0\ \forall j,k\in\{1,2,3\}\ j\neq k\nonumber, \text{ and \eqref{CDFV1B1}-\eqref{CDFV1B5}.} 
\end{align}
The linear program (\ref{OptProb}) can be solved using the simplex method~\cite{MatousekGartner} (or using the Fourier-Motzkin elimination). Solving \eqref{OptProb} while keeping (\ref{Ordering}) in mind, we obtain the following lower bound for the sum-capacity.
\begin{theorem}
\label{LowerBound:DF}
The sum-capacity of the Y-channel with condition \eqref{Ordering} satisfies $C_N\geq\underline{C}_{1}$ where
\begin{align}
\label{DFconstraint}
\underline{C}_{1}&=\min\left\{\frac{1}{2}\left[C(\min\{h_1^2,h_2^2+h_3^2\}P)+C(h_2^2P)+C(h_3^2P)\right],\right.\nonumber\\
&\hspace{1.2cm}\left.C\left(\left(h_1^2+h_2^2+h_3^2\right)P\right),2C\left(h_3^2P\right)\right\}.
\end{align}
\end{theorem}
\begin{IEEEproof}
The result can be obtained by solving the linear program in \eqref{OptProb} or by using  the Fourier-Motzkin elimination \cite{ElgamalKim} on the constraints of \eqref{OptProb}.
\end{IEEEproof}

Notice that as $P$ increases, the second argument in the $\min$ operation in $\underline{C}_{1}$ dominates the others. \alert{This follows since $$\lim_{\substack{P\to\infty}}\frac{\underline{C}_{1}}{\frac{1}{2}\log(P)}=\min\left\{\frac{3}{2},1,2\right\}=1,$$ dominated by the second term.} Thus, CDF achieves 1 DoF in the Y-channel.

\alert{By observing \eqref{CDFV1B1}-\eqref{CDFV1B5} we notice that the achievable rate of this scheme does not reduce to \eqref{2WRCCDF} unless if we set $h_3\to~\infty$, or if we remove the constraint \eqref{CDFV1B2} caused by decoding the unknown messages at user 3. The following is another CDF downlink variant which achieves \eqref{2WRCCDF} differently.}

\subsubsection{Downlink: Variant 2}
In this variant, after decoding the messages, the relay encodes each message $m_{jk}$ into an i.i.d. sequence $x_{r,jk}^n(m_{jk})$ where $X_{r,jk}\sim\mathcal{N}(0,P_{r,jk})$ such that $\sum_{j,k}P_{r,jk}\leq P$, for distinct $j,k\in\{1,2,3\}$. Then the relay sends the superposition of all these signals $$x_r^n=\sum_{j,k}x_{r,jk}^n.$$ 
Notice that user 3 knows $x_{r,12}^n$ and $x_{r,13}^n$. Thus, it can cancel the contribution of $x_{r,12}^n$ and $x_{r,13}^n$ from its received signal $y_1^n$, and then proceed to decode \emph{only its desired signals} $x_{r,21}^n$ and $x_{r,31}^n$ while treating the other signals as noise (contrary to the first variant where the undesired messages are also decoded). Thus, $R_{21}$ and $R_{31}$ must satisfy\footnote{\alert{The individual rate constraints $R_{21}\leq C(.)$ and $R_{31}\leq C(.)$ have been omitted since for the purpose of studying the sum-capacity, they are redundant given \eqref{CDFV2DL1}}}
\begin{align}
\label{CDFV2DL1}
R_{21}+R_{31}\leq C\left(\frac{h_1^2(P_{r,21}+P_{r,31})}{1+h_1^2(P_{r,23}+P_{r,32})}\right).
\end{align}
{A similar decoding strategy is used at the second and the third user, leading to}
\begin{align}
\label{CDFV2DL2}
R_{12}+R_{32}&\leq C\left(\frac{h_2^2(P_{r,12}+P_{r,32})}{1+h_2^2(P_{r,13}+P_{r,31})}\right),\\
\label{CDFV2DL3}
R_{13}+R_{23}&\leq C\left(\frac{h_3^2(P_{r,13}+P_{r,23})}{1+h_3^2(P_{r,21}+P_{r,12})}\right).
\end{align}

{By combining the bounds for the uplink \eqref{MACRC1}-\eqref{MACconstraint}, and those for the downlink \eqref{CDFV2DL1}-\eqref{CDFV2DL3}, we get the following achievable sum-capacity lower bound.}
\begin{theorem}
The sum-capacity of the Y-channel with condition \eqref{Ordering} satisfies
\begin{align}
\label{C2MaxProblem}
C_N\geq \underline{C}_2=\max_{\substack{P_{r,jk}\geq 0,\\ \sum_{j,k} P_{r,jk}\leq P}}\ \ \sum_{j=1}^3\sum_{\substack{k=1\\k\neq j}}R_{jk},
\end{align}
{such that the rates $R_{jk}$ satisfy \eqref{MACRC1}-\eqref{MACconstraint} and \eqref{CDFV2DL1}-\eqref{CDFV2DL3}.}
\end{theorem}

This maximization problem \eqref{C2MaxProblem} is tedious but can be carried out numerically. However, it can be easily seen that this scheme includes the CDF scheme of the BRC as a special case. Namely, set $P_{r,13}=P_{r,23}=P_{r,31}=P_{r,32}=0$ in \eqref{CDFV2DL1}-\eqref{CDFV2DL3}, and let $R_{13}=R_{23}=R_{31}=R_{32}=0$ to achieve \eqref{2WRCCDF}. Thus, as a corollary, we obtain the following.

\begin{corollary}
\label{LowerBound:CDF2}
{The sum-capacity lower bound $\underline{C}_{2}$ satisfies}
\begin{align}
\label{CDF2constraint}
\underline{C}_{2}&\geq\min\left\{C((h_1^2+h_2^2)P),2C\left(h_2^2P\right)\right\},
\end{align}
{where the right hand side of \eqref{CDF2constraint} is the achievable sum-rate of \eqref{2WRCCDF}.}
\end{corollary}
Keep in mind that this variant of CDF, variant 2, achieves higher rates than \eqref{CDF2constraint} in general, namely, it achieves \eqref{C2MaxProblem} where the rates $R_{13}$, $R_{23}$, $R_{31}$, and $R_{32}$ are not necessarily zero.

\subsection{Functional Decode-and-Forward}
\label{LowerBound2}
\begin{figure*}[th]
\centering
\includegraphics[width=0.8\textwidth]{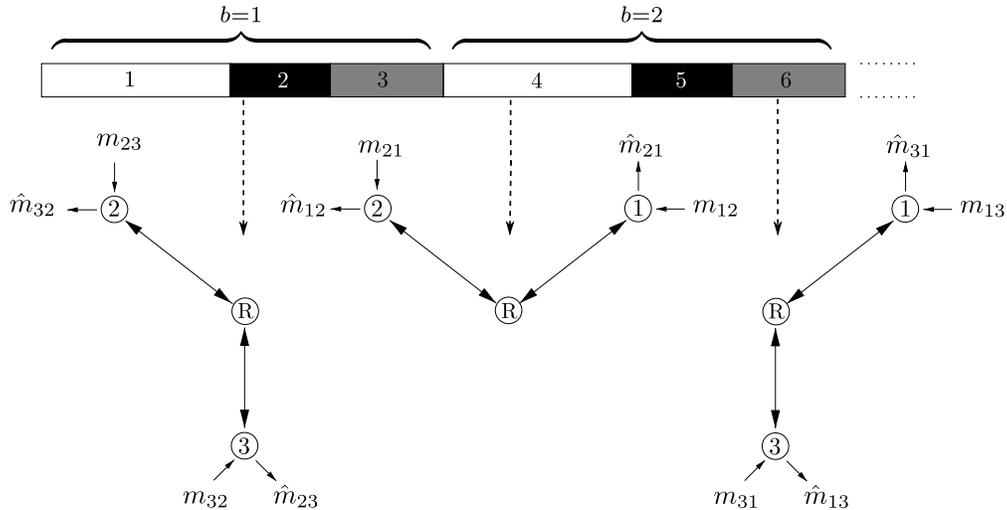}
\caption{An illustration of the FDF scheme showing blocks 1 and 2. The procedure in slots with the same color is the same, i.e., the slots are repeated with a period of 3 slots.}
\label{Fig:Frames}
\end{figure*}

\alert{In this section, we describe functional decode-and-forward (FDF), another scheme for the Y-channel based on compute-and-forward. In the proposed scheme, a transmission block is divided into 3 time slots, where only 2 users are active in each slot as shown in Fig. \ref{Fig:Frames}. These slots will be indexed as slot $3b+s$ where $b=0,1,2,\dots$ is the block index and $s\in\{1,2,3\}$ is the slot index. The duration of slot 1 where users 1 and 2 are active is $\alpha_{12}n$ symbols. Slot 2 where users 2 and 3 are active, and slot 3 where users 3 and 1 are active, have $\alpha_{23}n$ and $\alpha_{31}n$ symbols, respectively, where $\alpha_{12}+\alpha_{23}+\alpha_{31}=1$. The 3 slots are repeated periodically where the procedure in slot $3b+s$ is the same as that in slot $s$.} 

In each slot, the 2 active users and the relay use network coding with lattice codes as in the BRC \cite{NarayananPravinSprintson}. Briefly, in slot $3b+1$, the two active users (1 and 2) send $\alpha_{12}n$ symbols of their signals $x_1^n(m_{12}(b))$ and $x_2^n(m_{21}(b))$ to the relay, where $m_{12}(b),m_{21}(b)\in\{1,\dots,2^{nR_{12}}\}$\footnote{Notice that we fixed $R_{12}=R_{21}$. One could also set an asymmetric rate allocation by using the scheme in \cite{NamChungLee_IT}.}. The relay, after collecting $n$ symbols, decodes the superposition of $x_1^n(m_{12}(b))$ and $x_2^n(m_{21}(b))$, maps it to $u_{12}(b)\in\{1,\dots,2^{nR_{12}}\}$ and sends $x_r^n(u_{12}(b))$. The receiver at user 1 for instance, after collecting $n$ symbols, decodes $m_{21}$ from the its received signal $y_1^n$ using its message $m_{12}$ as side information. The details of this scheme are given in Appendix \ref{FDF}. The achievable sum-rate is given in the following Theorem.

\begin{theorem}
\label{LowerBound:FDF}
{The sum-capacity of the Y-channel with condition \eqref{Ordering} satisfies $C_N\geq\underline{C}_{3}$ where}
\begin{align*}
\underline{C}_{3}&=2\alpha_{12}\min\left\{C^+\left(h_2^2P_{21}^*-\frac{1}{2}\right),C\left(h_2^2P\right)\right\}\\
&+2\alpha_{23}\min\left\{C^+\left(h_3^2P_{31}^*-\frac{1}{2}\right),C\left(h_3^2P\right)\right\}\\
&+2\alpha_{31}\min\left\{C^+\left(h_3^2P_{32}^*-\frac{1}{2}\right),C\left(h_3^2P\right)\right\},
\end{align*}
maximized over $\alpha_{12},\alpha_{23},\alpha_{31}\in[0,1]$ with $\alpha_{12}+\alpha_{23}+\alpha_{31}=1$ and 
\begin{align}
\label{P21*}
h_2^2P_{21}^*&=\min\left\{\frac{h_1^2P}{\alpha_{12}+\alpha_{31}},\frac{h_2^2P}{\alpha_{12}+\alpha_{23}}\right\},\\
\label{P31*}
h_3^2P_{31}^*&=\min\left\{\frac{h_1^2P}{\alpha_{12}+\alpha_{31}},\frac{h_3^2P}{\alpha_{23}+\alpha_{31}}\right\},\\
\label{P32*}
h_3^2P_{32}^*&=\min\left\{\frac{h_2^2P}{\alpha_{12}+\alpha_{23}},\frac{h_3^2P}{\alpha_{23}+\alpha_{31}}\right\}.
\end{align}
\end{theorem}

It can easily be seen that FDF achieves a DoF of 2, and thus achieves the DoF of the Y-channel. This scheme clearly outperforms CDF at high $P$. However, at low $P$, namely, when $h_2^2P_{21}^*, h_3^2P_{31}^*, h_2^2P_{32}^*\leq \frac{1}{2}$, this scheme achieves zero rate. In this regime, the CDF scheme performs better and is thus useful.

Note that the FDF scheme is designed to serve all three users in a time sharing fashion. Since in this scheme the users are not active in general all the time, they can transmit at a power larger than $P$ (\eqref{P21*}-\eqref{P32*}) while still satisfying the average power constraint. A simple lower bound on $\underline{C}_{3}$ can be found by setting $\alpha_{12}=1$, i.e., switching user 3 completely off, and operating the Y-channel as a BRC with FDF where only the two strongest users communicate\footnote{Note that this is not the optimal time-sharing in general since it does not exploit the available transmit power of user 3.}. The scheme in this case reduces to the BRC FDF scheme \cite{NamChungLee}. This leads to the following corollary.
\begin{corollary}
\label{LowerBound:FDF_2User}
{The sum-capacity lower bound $\underline{C}_3$ satisfies}
\begin{align*}
\underline{C}_{3}&\geq2\min\left\{C^+\left(h_2^2P-\frac{1}{2}\right),C\left(h_2^2P\right)\right\}.
\end{align*}
\end{corollary}

Finally, we would like to note that the given scheme does not use the full power at all users (due to the alignment constraint, cf. \eqref{PowerAlignment}). To achieve higher rates, the users can use the remaining power to send an extra codeword superimposed on top of the lattice codeword as in \cite{SezginAvestimehrKhajehnejadHassibi}.

\subsection{Evaluation}
Figure \ref{sum_rate_asymmetric} shows a plot of the obtained upper and lower bounds versus $\mathsf{SNR}$. Namely, the plotted bounds are: 
\begin{figure*}[t]
\centering
\includegraphics[width=.9\textwidth]{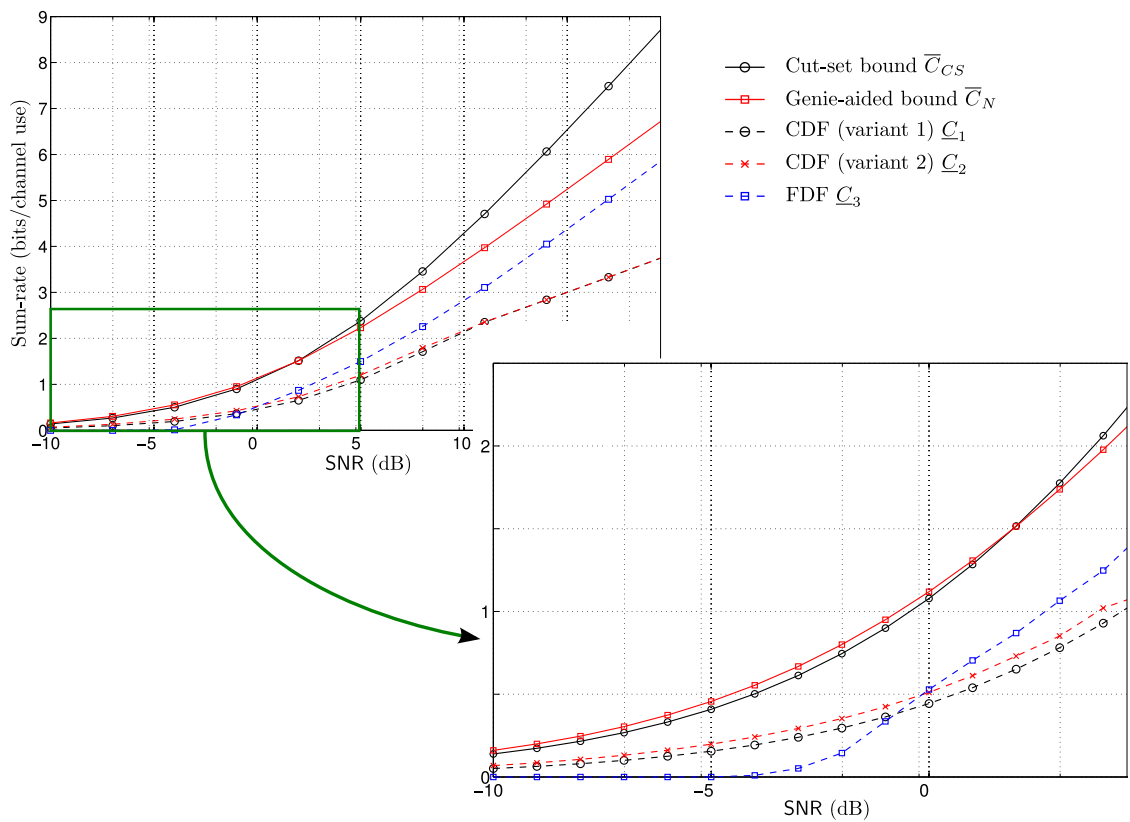}
\caption{A plot of the upper and lower bounds for a Y-channel with $h_1=1$, $h_2=0.8$, and $h_3=0.6$.}
\label{sum_rate_asymmetric}
\end{figure*}
\begin{itemize}
\item the upper bound obtained with the cut-set approach $\overline{C}_{CS}$, 
\item the upper bound obtained with the genie-aided approach $\overline{C}_{N}$, 
\item the complete decode-and-forward lower bounds $\underline{C}_{1}$ and $\underline{C}_{2}$, and 
\item the functional decode-and-forward lower bound $\underline{C}_{3}$, 
\end{itemize}
for a Y-channel with $h_1=1$, $h_2=0.8$, $h_3=0.6$.

It can be seen that $\overline{C}_{N}$ is tighter than $\overline{C}_{CS}$ at moderate to high $\mathsf{SNR}$. It can also be seen that the gap between $\overline{C}_{N}$ and $\underline{C}_{3}$ becomes constant as $\mathsf{SNR}$ increases, which reflects the DoF of the channel. In the following section, we characterize this constant gap.

\section{Bounding the Gap between the Upper and Lower Bounds}
\label{GapCalculation:General}

The derived upper and lower bounds are within a constant additive and multiplicative gap as we show next. We calculate these gaps in order to provide an approximate characterization of the sum-capacity of the Y-channel.

\subsection{Multiplicative Gap Calculation}
We bound the multiplicative gap first. That is, we bound the ratio of the upper bound to the lower bound. Using Corollary \ref{LowerBound:CDF2}, we can always write
\begin{align}
\label{BBCDF2}
C_N&\geq\underline{C}_{2}\geq C\left(h_2^2P\right).
\end{align}
Moreover, from the bound $\overline{C}_{CS}$, we have
\begin{align}
C_N&\leq\overline{C}_{CS}\nonumber\\
&\leq C\left((h_2^2+h_3^2)P\right)+C\left(h_2^2P\right)+C\left(h_3^2P\right)\nonumber\\
&\stackrel{(a)}{\leq} 2C\left(h_2^2P\right)+2C\left(h_3^2P\right)\nonumber\\
\label{BBCS}
&\leq 4C\left(h_2^2P\right),
\end{align}
where $(a)$ follows from the concavity of $\log(1+x)$. Therefore
$\overline{C}_{CS}\leq 4\underline{C}_{2}$, leading to a multiplicative gap of 4:
\begin{align*}
\frac{\overline{C}_{CS}}{4}\leq C_N\leq \overline{C}_{CS}.
\end{align*}

\subsection{Additive Gap}
Now we calculate the additive gap, which we split into two cases: $h_2^2P\leq\frac{1}{2}$ and $h_2^2P>\frac{1}{2}$.

\subsubsection{Case $h_2^2P\leq\frac{1}{2}$}
Consider the lower bound $\underline{C}_{2}$ and the upper bound $\overline{C}_{CS}$. These bounds can be used to obtain the following
\begin{align*}
\overline{C}_{CS}-\underline{C}_{2}&\stackrel{(b)}{\leq}3C(h_2^2P)\stackrel{(c)}{\leq} \frac{3}{2}\log\left(\frac{3}{2}\right),
\end{align*}
where $(b)$ follows by using \eqref{BBCDF2} and \eqref{BBCS}, and $(c)$ follows since $h_2^2P\leq\frac{1}{2}$.

\subsubsection{Case $h_2^2P>\frac{1}{2}$}
\label{3UserGap}
Using Corollary \ref{LowerBound:FDF_2User} with $h_2^2P>\frac{1}{2}$ we can write
\begin{align*}
C_N&\geq\underline{C}_{3}\geq2C\left(h_2^2P-\frac{1}{2}\right).
\end{align*}
Using this we bound the gap between the upper bound $\overline{C}_{N}$ and the lower bound $\underline{C}_{3}$ as follows
\begin{align*}
\overline{C}_{N}-\underline{C}_{3}&\leq C(h_2^2P+h_3^2P)+C((|h_2|+|h_3|)^2P)\\
&\quad-2C\left(h_2^2P-\frac{1}{2}\right)\\
&\leq 2C(4h_2^2P)-2C\left(h_2^2P-\frac{1}{2}\right)\\
&\leq 2C(2h_2^2P)+1-2C\left(h_2^2P-\frac{1}{2}\right)\\
&= 2.
\end{align*}

\begin{figure}[t]
\centering
\psfragscanon
\psfrag{x}[t]{$h_2$}
\psfrag{y}[b]{$h_3$}
\includegraphics[width=.5\columnwidth]{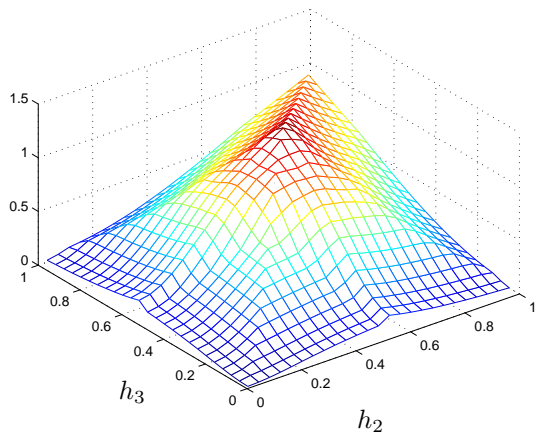}
\caption{The gap between the upper and lower bounds plotted as a function of $h_2$ and $h_3$ for a Y-channel with $\mathsf{SNR}=10$ dB and $h_1=1$. Notice that the gap is always smaller than 1 in this case.}
\label{Fig:Gap}
\end{figure}

To summarize, combining the additive and the multiplicative gaps we can write 
\begin{align*}
\max\left\{\overline{C}_{CS}-\frac{3}{2}\log\left(\frac{3}{2}\right),\frac{\overline{C}_{CS}}{4}\right\}\leq C_N\leq \overline{C}_{CS}
\end{align*}
if $h_2^2P\leq\frac{1}{2}$, and 
\begin{align*}
\max\left\{\overline{C}_{N}-2,\frac{\overline{C}_{CS}}{4}\right\}\leq C_N\leq \min\{\overline{C}_{N},\overline{C}_{CS}\}
\end{align*}
otherwise. Thus, we have bounded the gap between our sum-capacity upper and lower bounds by a constant independent of the channel coefficients. As a result, noting that $\frac{3}{2}\log\left(\frac{3}{2}\right)<2$, {\it we have characterized the sum-capacity of the Y-channel within an additive gap of at most 2 bits and a multiplicative gap of 4 for the non-restricted Y-channel for any value of $P$}. Notice that the multiplicative gap is important for the case of low power, especially when the additive gap itself becomes larger than the upper bound and thus obsolete. The gap between the upper and lower bounds is plotted as a function of $h_2$ and $h_3$ for a Y-channel with $P=10$ and $h_1=1$ in Fig. \ref{Fig:Gap}.

Let us now consider the symmetric Y-channel where $h_1=h_2=h_3$. In this case, we can show that the gap between the upper and lower bounds is always less than 1 bit.

\subsection{Gap Calculation for the Symmetric Y-channel}
\begin{figure}[t]
\centering
\psfragscanon
\psfrag{x}[t]{$\mathsf{SNR}$(dB)}
\psfrag{y}[b]{Sum Rate}
\psfrag{UB}[bl]{\footnotesize Upper Bound: $\min\{\overline{C}_{CS},\overline{C}_{N}\}$}
\psfrag{LB}[Bl]{\footnotesize Lower Bound: $\max\{\underline{C}_{1},\underline{C}_{2},\underline{C}_{3}\}$}
\includegraphics[width=.5\columnwidth]{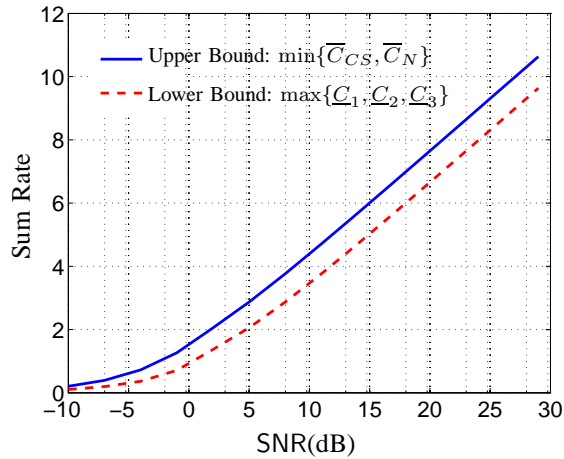}
\caption{A plot of the upper and lower bounds for the symmetric Y-channel.}
\label{Bounds_R}
\end{figure}

\alert{In the symmetric Y-channel, $h_1=h_2=h_3=h$. Moreover, we can assume that $h=1$ without loss of generality}. In this case, we can rewrite the bounds we have in a simpler form. Starting from Corollary \ref{CSG}, for the symmetric Y-channel we have
\begin{align*}
C_N\leq\overline{C}_{CS}=3C(P).
\end{align*}
Moreover, from Theorem \ref{SRUBG} we have the upper bound
\begin{align}
\label{BCs}
C_N&\leq\overline{C}_{N}=2C(2P),
\end{align}
and from Theorems \ref{LowerBound:DF}, \ref{LowerBound:FDF_2User}, and \ref{LowerBound:FDF}, we have the lower bounds
\begin{align}
C_N\geq\underline{C}_{1}&=\min\left\{C(3P),\frac{3}{2}C(P)\right\}\\
\label{BCii}
C_N\geq\underline{C}_{2}&\geq\min\left\{C\left(2P\right),2C(P)\right\}\\
\label{BCiii}
C_N\geq\underline{C}_{3}&\geq2\min\left\{C^+\left(P-\frac{1}{2}\right),C(P)\right\}.
\end{align}
Now that we have upper and lower bounds for the sum-capacity of the symmetric Y-channel, we can upper bound the gap between them by 1 bit,
\begin{align}
\label{Dg}
\min\{\overline{C}_{CS},\overline{C}_{N}\}-\max\{\underline{C}_{1},\underline{C}_{2},\underline{C}_{3}\}\leq 1,
\end{align}
for any value of $P$. Figure \ref{Bounds_R} shows the upper and lower bounds for a symmetric Y-channel, where it can be seen that the gap is always less than 1 bit.

\section{Upper Bounds for the Restricted Y-channel}
\label{Section:RYC}
In this section, we impose an additional constraint on the Y-channel. That is, we consider the Y-channel with a restricted encoder (\ref{SourceEncoder_R}). Recall that the difference between the non-restricted Y-channel and the restricted one is that the transmit signals can be dependent in the former while they are independent in the later. 
The independence of the transmit signals can lead to a tighter upper bound. Namely, the upper bound in Theorem \ref{SRUBG} can be tightened leading to a smaller gap to the lower bound. We start with the following lemma. 

\begin{lemma}
\label{RUB}
The achievable rates in the restricted Y-channel must satisfy
\begin{align*}
R_{kj}+R_{lj}+R_{kl}&\leq C((h_k^2+h_l^2)P)
\end{align*}
for all distinct $j,k,l\in\{1,2,3\}$.
\end{lemma}
\begin{IEEEproof}
The proof is similar to that of Lemma \ref{GUB}, except that in this case, the random variables representing the transmit signals, $X_1$, $X_2$ and $X_3$, are independent. Details are given in Appendix \ref{ProofRestricted}.
\end{IEEEproof}
\alert{Notice the difference between the bounds in Lemmas \ref{GUB} and \ref{RUB}, which is mainly due to the dependence of the transmit signals of the users in the non-restricted Y-channel and their independence in the restricted one.} Combining Lemma \ref{FromRelay} and \ref{RUB} we get for the restricted Y-channel
\begin{align}
R_{kj}+R_{lj}+R_{kl}&\leq \min\left\{C((h_j^2+h_l^2)P),C((h_k^2+h_l^2)P)\right\},
\label{3BoundR}
\end{align}
from which we have the following theorem, which provides a tighter upper bound than $\overline{C}_N$.
\begin{theorem}
\label{SRUBR}
The sum-capacity of the restricted Y-channel is upper bounded by $\overline{C}_{R}$,
\begin{align}
\label{RSCUB}
C_R\leq\overline{C}_{R}&=2C(h_2^2P+h_3^2P).
\end{align}
\end{theorem}
\begin{IEEEproof}
By evaluating (\ref{3BoundR}) for $(j,k,l)=(1,2,3)$, and for $(j,k,l)=(2,1,3)$ and  adding the two obtained bounds, we obtain the desired result.
\end{IEEEproof}

\subsection{Gap Calculation}
Since $C_R\leq C_N$, all upper bounds for the non-restricted Y-channel continue to hold for the restricted one. However, we need not to consider $\overline{C}_{N}$ anymore since $\overline{C}_{R}$ in (\ref{RSCUB}) is clearly tighter. The developed lower bounds also continue to hold since the corresponding transmission schemes used restricted encoders. 

While all calculated gaps hold true, the gap for $h_2^2P>\frac{1}{2}$ can be made tighter
\begin{align*}
\overline{C}_{R}-\underline{C}_{3}&\leq2C(2h_2^2P)-2C\left(2h_2^2P\right)+1= 1,
\end{align*}
where we used $h_3^2\leq h_2^2$ (\ref{Ordering}). As a result, for $h_2^2P\leq\frac{1}{2}$ we have
\begin{align*}
\max\left\{\overline{C}_{CS}-\frac{3}{2}\log\left(\frac{3}{2}\right),\frac{\overline{C}_{CS}}{4}\right\}&\leq C_R\leq\overline{C}_{CS}.
\end{align*}
and for $h_2^2P>\frac{1}{2}$ we have
\begin{align*}
\max\left\{\overline{C}_{R}-1,\frac{\overline{C}_{CS}}{4}\right\}\leq C_R\leq \min\{\overline{C}_{R},\overline{C}_{CS}\}.
\end{align*}
Since $\frac{3}{2}\log\left(\frac{3}{2}\right)<1$, then for the restricted Y-channel we obtain an additive gap of 1 bit, and a multiplicative gap of 4. For the symmetric restricted Y-channel, the same gap of 1 bit holds as that in the asymmetric one.

\section{\alert{Extensions}}
\label{Sec:Extensions}
In this section, we extend the sum-capacity approximation of the Y-channel within a constant gap to the network with more than 3 users. We obtain the sum-capacity of the $K$-user case within a constant gap. Moreover, we discuss the non-reciprocal case briefly, where we show that the same schemes used for constant gap characterization in the reciprocal case are not suitable for the non-reciprocal case. Namely, these lead to a gap between the upper and lower bounds that is dependent on the channel parameters as we shall see in Section \ref{Sec:NonReciprocal}.

\subsection{$K$ Users}
\label{K-user-Star}
The upper bounds developed in Lemmas \ref{FromRelay} and \ref{GUB} can be extended to the larger network with $K$ users and one relay, i.e., a $K$-user star channel with full message exchange. This extension leads to the sum-capacity of the $K$-user star channel, which is stated in the following theorem.
\begin{theorem}
\label{Thm:KuserCap}
The sum-capacity of the $K$-user star channel, denoted by $C_{\Sigma}$, satisfies
\begin{align*}
\overline{C}_{\Sigma}-2\log(K-1)\leq C_{\Sigma}\leq\overline{C}_{\Sigma},
\end{align*}
where
\begin{align}
\label{KUserUB}
\overline{C}_\Sigma= \frac{1}{2}\log(1+(\|\vec{h}\|_2)^2P)+\frac{1}{2}\log\left(1+(\|\vec{h}\|_1)^2P\right).
\end{align}
with $\vec{h}=(h_2,h_3,\cdots,h_K)$, and $\|\vec{h}\|_1$ and $\|\vec{h}\|_2$ are its $\ell^1$-norm and $\ell^2$-norm, respectively.
\end{theorem}
\begin{IEEEproof}
The converse of this theorem is provided by a genie-aided bound. Recall that the key for developing the genie-aided upper bounds of the Y-channel was a careful design of a genie-aided channel where some bounds can be combined using the chain rule of mutual-information. In the 3-user case (Y-channel), we had to add up two bounds to obtain an upper bound on the sum of 3 components of $\vec{R}$. In the $K$ user case, we can use $K-1$ bounds, which when added produce an upper bound on the sum of $K(K-1)/2$ components of $\vec{R}$ (which is $K(K-1)$ dimensional in this case). Namely, we can obtain the following upper bound
\begin{align*}
n\sum_{k=2}^{K}\sum_{j=1}^{k-1}R_{jk}\leq I\left(X_r^n;Y_K^n,\dots,Y_{2}^n\right)+n\varepsilon_n,
\end{align*}
(see details in Appendix \ref{Proof:KUserStar}). This bound resembles the capacity of a SIMO point-to-point channel which can be written as \cite{TseViswanath}
\begin{align}
\label{BoundAll1G}
\sum_{k=2}^{K}\sum_{j=1}^{k-1}R_{jk}&\leq \frac{1}{2}\log(1+(\|\vec{h}\|_2)^2P).
\end{align}
We can also obtain the bound
\begin{align}
\label{BoundAll2G}
\sum_{k=1}^{K-1}\sum_{j=k+1}^{K}R_{j,k}&\leq \frac{1}{2}\log\left(1+(\|\vec{h}\|_1)^2P\right),
\end{align}
using a similar derivation as for the 3-user case in Appendix \ref{GeneralProof} (see Appendix \ref{Proof:KUserStar}). By adding the 2 bounds, \eqref{BoundAll1G} and \eqref{BoundAll2G}, we obtain the upper bound given by \eqref{KUserUB}. The sum-capacity lower bound is given by the BRC scheme, i.e., $\underline{C}_3$ in Corollary \ref{LowerBound:FDF_2User}. Finally, using some simple steps, we can show that, given $h_2^2P>\frac{1}{2}$, the gap between the achievable sum-rate using this scheme and the derived upper bound is less than $2\log\left(K-1\right)$ (see Appendix \ref{Proof:KUserStar_Gap} for more details) which completes the proof.
\end{IEEEproof}

Having derived the sum-capacity of the $K$-user case within a constant gap, we can notice that this sum-capacity has the same behavior as for the BRC and the Y-channel, i.e., the sum capacity behaves as $\log(P)+o(\log(P))$. Thus, the $K$-user star channel has the same DoF of 2 for all $K\geq2$. Notice also that the gap given in Theorem \ref{Thm:KuserCap} recovers the gap of 2 bits obtained in Section \ref{3UserGap} for the 3-user case.

\subsection{\alert{Non-reciprocal Y-channel}}
\label{Sec:NonReciprocal}
In the paper, we have assumed that the channels are reciprocal. Thus, the results and especially the constant gap holds given that the channels are reciprocal. While the characterization of the sum-capacity of the non-reciprocal case is beyond the scope of this paper, it's worth to briefly address the impact of non-reciprocal channels on the gap analysis. As the derived upper bounds apply for the reciprocal Y-channel, the bounds in Lemmas \ref{FromRelay} and \ref{GUB} have to be rewritten for the non-reciprocal case as follows
\begin{align}
\label{NRBound1}
R_{kj}+R_{lj}+R_{kl}&\leq C((h_{rj}^2+h_{rl}^2)P),\\
\label{NRBound2}
R_{kj}+R_{lj}+R_{kl}&\leq C((|h_{kr}|+|h_{lr}|)^2P),
\end{align}
where $h_{rj}$ and $h_{jr}$ denote the channel from the relay to user $j$ and vice versa, respectively. Furthermore, the transmission schemes can be also applied for the non-reciprocal Y-channel after the necessary modifications. For instance, in the CDF scheme, the channels $h_1$, $h_2$ and $h_3$ have to be replaced by $h_{1r}$, $h_{2r}$ and $h_{3r}$, respectively in the rate constraints \eqref{MACRC1}-\eqref{MACconstraint}, and by $h_{r1}$, $h_{r2}$, and $h_{r3}$, respectively in the rate constraints \eqref{DFC1}-\eqref{DFC3}. Then, the achievable sum-rate can be derived by solving a linear program as that in\footnote{Since there is no general ordering of the strengths of the channels in the non-reciprocal case, all the rate constraints of CDF have to be taken into account in the linear program.} \eqref{OptProb}. This applies for both variants of CDF. On the other hand, the simple FDF scheme leading to the lower bound in Corollary \ref{LowerBound:FDF_2User}, leads to the following lower bound when applied to the non-reciprocal case
\begin{align*}
2\min\left\{C^+\left(\min\{h_{jr}^2,h_{kr}^2\}P-\frac{1}{2}\right),C\left(\min\{h_{rj}^2,h_{rk}^2\}P\right)\right\},
\end{align*}
with $j,k\in\{1,2,3\}$. Note that for the non-reciprocal case, since no general ordering of the channel exists, users 1 and 2 are not necessarily the two strongest users anymore. As shown in the following example, these bounds can be used to obtain a sum-capacity characterization within a gap which is independent of the transmit power but dependent on the channel coefficients.
\begin{claim}
\alert{The sum-capacity of the non-reciprocal Y-channel can not be characterized within a universal constant gap by using the developed upper and lower bounds in Sections \ref{UpperBounds} and \ref{LowerBounds}.}
\end{claim}
\alert{To prove this claim, we use a special example, and show that for this example the difference between the bounds is a function of the channel parameters. The existence of such an example shows that in general, a universal constant gap does not hold for the non-reciprocal Y-channel using our bounds, and hence proves the claim.}

\subsubsection{Example}
Consider a Y-channel with 
\begin{align}
\label{NROrdering}
h_{1r}=h_{r1}>h_{2r}>h_{r2}>h_{3r}=h_{r3}. 
\end{align}
In this case, in order to get the best sum-capacity upper bound given by \eqref{NRBound1} and \eqref{NRBound2}, we need to know more about the relations between these channel coefficients. However, we can relax these bounds while still preserving their behavior (within a constant gap) as follow. Set $(j,k,l)=(2,1,3)$ in \eqref{NRBound1} and \eqref{NRBound2} to obtain
\begin{align}
\label{NRNonRelaxedBound}
&\hspace{-.5cm}R_{12}+R_{13}+R_{32}\nonumber\\
&\leq \min\{C((h_{r2}^2+h_{r3}^2)P),C((|h_{1r}|+|h_{3r}|)^2P)\}\\
&\leq \min\{C(4h_{r2}^2P),C(4h_{1r}^2P)\}\\
\label{NRRelaxedBound}
&=C(4h_{r2}^2P).
\end{align}
Notice that \eqref{NRRelaxedBound} is within a constant of \eqref{NRNonRelaxedBound}. Similarly, we can relax the bounds \eqref{NRBound1} and \eqref{NRBound2} for different $(j,k,l)$. Then, by maximizing the sum-rate subject to these bounds, we can get the sum-capacity upper bound
\begin{align*}
C_\Sigma\leq C(4h_{r2}^2P)+C(4h_{2r}^2P).
\end{align*}
Now, the lower bound obtained by using the FDF scheme is given by (Corollary \ref{LowerBound:FDF_2User})
\begin{align*}
C_\Sigma\geq 2\min\left\{C^+\left(h_{2r}^2P-\frac{1}{2}\right),C\left(h_{r2}^2P\right)\right\}.
\end{align*}
By comparing the upper and lower bounds, we can use similar steps as those in Section \ref{3UserGap} (given $h_{2r}^2P>\frac{1}{2}$) to bound the gap between the bounds by
\begin{align*}
\frac{1}{2}\max\left\{5+\log\left(\frac{h_{r2}^2}{h_{2r}^2}\right),3+\log\left(\frac{h_{2r}^2}{h_{r2}^2}\right)\right\}.
\end{align*}
\begin{figure}[t]
\centering
\psfragscanon
\psfrag{x}[t]{$h_{r2}$}
\psfrag{y}[b]{$h_{2r}$}
\includegraphics[width=.5\columnwidth]{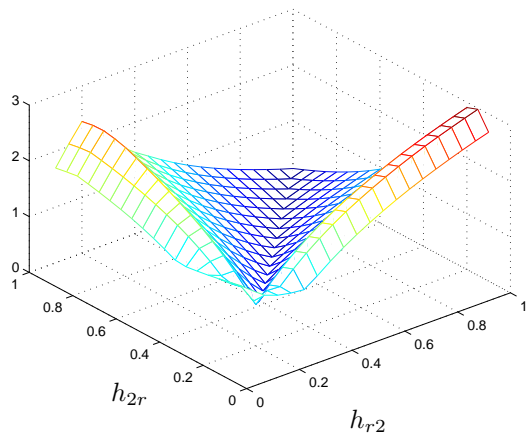}
\caption{The gap between the upper and lower bounds for a non-reciprocal Y-channel with $h_{1r}=h_{r1}=1$, $h_{3r}=h_{r3}=0.1$, plotted as a function of $h_{2r}$ and $h_{r2}$ for a Y-channel with $\mathsf{SNR}=30$ dB. Notice that the gap in this case can be larger than 2 (the maximum gap for the reciprocal case).}
\label{Fig:Gap_NR}
\end{figure}
As can be seen in this example, the derived gap is a function of the channel gains in contrast to the reciprocal case, and thus, can not be bounded by a universal constant. Although this gap does not depend on $\mathsf{SNR}$, it can be arbitrarily large depending on the channels (see Fig. \ref{Fig:Gap_NR}). Therefore, the simple FDF scheme we used for the reciprocal Y-channel is not enough for characterizing the sum-capacity of the non-reciprocal one within a universal constant. In future work, one might investigate whether more sophisticated schemes such as a superposition of nested-lattice codes (for bidirectional communication) and Gaussian random codes (for uni-directional communication) \cite{ChaabanSezginISIT, SezginAvestimehrKhajehnejadHassibi}, or a combination of quantize-and-forward and decode-and-forward \cite{AvestimehrSezginTse} or other schemes give a universal constant gap.

\section{Conclusion}
\label{Conclusion}
\alert{We have shown that for the Y-channel, a setup with three users and one relay where each user sends 2 messages, one to each other user via the relay, the total DoF is 2. This is the same DoF of the bi-directional relay channel. While the achievability of this DoF is straightforward (using the same scheme as for the bi-directional relay channel), the converse is less obvious. We have developed a new upper bound which shows this result, and hence characterizes the DoF of the channel. Moreover, the gap between the bounds is evaluated. We have shown an additive gap of 2 bits and a multiplicative gap of 4 for the non-restricted Y-channel, and an additive gap of 1 bit and a multiplicative gap of 4 for the restricted one.}

\alert{The DoF is not characterized by the cut-set bounds, contrary to other setups and contrary to the MIMO Y-channel (under some condition on the number of antennas). Furthermore, we have extended this result to the $K$-user case and showed that the channel has the same DoF as the bi-directional relay channel, i.e., 2 DoF for all $K\geq3$. The strategy used for the BRC achieves our new sum-capacity upper bound within a constant gap which is independent of the power and only depends on the number of users in the system.}

An interesting future direction is to study bi-directional communication in the case that there are multiple relays available to help. For the case of one-way traffic, recent results for two-unicast networks~\cite{GouJafarJeonChung,ShomoronyAvestimehr,ShomoronyAvestimehr_IT} and multi-unicast networks~\cite{ShomoronyAvestimehr_Allerton} demonstrate that relays can provide significant capacity gains by also assisting with interference management. Understanding the limits of bi-directional communication via interfering relays (even from degrees-of-freedom perspective) is still an open problem.

\begin{appendices}

\section{Proof of Corollary \ref{CSG}}
\label{CSGProof}
From the first cut-set bound (\ref{CS1}), we have
\begin{align}
R_{jk}+R_{jl}&\leq I(X_j;Y_r|X_k,X_l,X_r)\nonumber\\
&=h(Y_r|X_k,X_l,X_r)-h(Z_r)\nonumber\\
&\leq h(h_jX_j+Z_r)-h(Z_r)\nonumber\\
\label{FCSB1}
&\leq C(h_j^2P),
\end{align}
and
\begin{align}
R_{jk}+R_{jl}&\leq I(X_r;Y_k,Y_l|X_k,X_l)\nonumber\\
&=h(Y_k,Y_l|X_k,X_l)-h(Y_k,Y_l|X_k,X_l,X_r)\nonumber\\
&\leq h(Y_k,Y_l)-h(Z_k,Z_l)\nonumber\\
\label{FCSB2}
&\leq C(h_k^2P+h_l^2P),
\end{align}
where we have used the Gaussian distribution to maximize these bounds. From (\ref{FCSB1}) and (\ref{FCSB2}) we obtain (\ref{CSG1}). Using (\ref{CS2}), we have
\begin{align}
R_{jl}+R_{kl}&\leq I(X_j,X_k;Y_r|X_l,X_r)\nonumber\\
&= h(Y_r|X_l,X_r)-h(Y_r|X_l,X_r,X_j,X_k)\nonumber\\
&\leq h(h_jX_j+h_kX_k+Z_r)-h(Z_r)\nonumber\\
&\leq C(h_j^2P+h_k^2P+2h_jh_k\rho_{jk}P)\nonumber\\
\label{SCSB1}
&\leq C((|h_j|+|h_k|)^2P),
\end{align}
where $\rho_{jk}=\mathbb{E}[X_jX_k]/P\in[-1,1]$, and
\begin{align}
R_{jl}+R_{kl}&\leq I(X_r;Y_l|X_l)\nonumber\\
&= h(Y_l|X_l)-h(Y_l|X_l,X_r)\nonumber\\
&\leq h(Y_l)-h(Z_l)\nonumber\\
\label{SCSB2}
&\leq C(h_l^2P).
\end{align}
From (\ref{SCSB1}) and (\ref{SCSB2}) we obtain (\ref{CSG2}).

\section{\alert{Proof of Lemma \ref{FromRelay}}}
\label{ProofFromRelay}
\begin{figure}[t]
\centering
\includegraphics[width=.5\columnwidth]{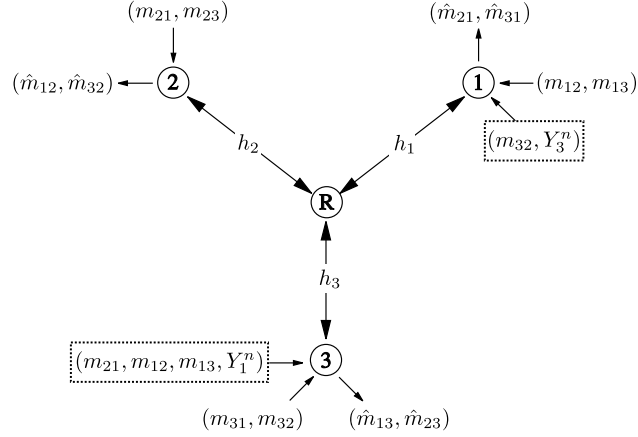}
\caption{The Y-channel with side information}
\label{GAYC_FR}
\end{figure}

{Here, we prove Lemma \ref{FromRelay} with $(j,k,l)=(1,2,3)$, all other cases follow similarly.} By giving $(Y_3^n,m_{32})$ to receiver 1, and $(Y_1^n,m_{21},m_{12},m_{13})$ to receiver 3 as additional information as shown in Figure \ref{GAYC_FR}, and using Fano's inequality, we have
\begin{align}
n(R_{21}+R_{31}-\epsilon_{1n})&\leq I(m_{21},m_{31};Y_1^n,m_{12},m_{13},Y_3^n,m_{32})\nonumber\\
\label{FP}
&= I(m_{21},m_{31};Y_1^n,Y_3^n|m_{12},m_{13},m_{32}),
\end{align}
and
\begin{align}
n(R_{23}-\epsilon_{2n})&\leq I(m_{23};Y_3^n,m_{31},m_{32},Y_1^n,m_{21},m_{12},m_{13})\nonumber\\
\label{SP}
&= I(m_{23};Y_1^n,Y_3^n|m_{31},m_{32},m_{21},m_{12},m_{13}),
\end{align}
where $\epsilon_{1n},\epsilon_{2n}\to0$ as $n\to\infty$, where (\ref{FP}) and (\ref{SP}) follow by using the chain rule and from the independence of the messages. Adding (\ref{FP}) and (\ref{SP}) and using the chain rule of mutual information, we get
\begin{align*}
&n(R_{21}+R_{31}+R_{23}-\epsilon_{n})\\
&\quad\leq I(m_{21},m_{31},m_{23};Y_1^n,Y_3^n|m_{12},m_{13},m_{32}),
\end{align*}
where $\epsilon_{n}=\epsilon_{1n}+\epsilon_{2n}\to0$ as $n\to\infty$. We continue
\begin{align*}
n(R_{21}+R_{31}+R_{23}-\epsilon_{n})&\stackrel{(a)}{\leq} h(Y_1^n,Y_3^n)-h(Y_1^n,Y_3^n|\vec{m})\\
&\stackrel{(b)}{\leq} h(Y_1^n,Y_3^n)-h(Y_1^n,Y_3^n|\vec{m},X_r^n)\\
&\stackrel{(c)}{=} h(Y_1^n,Y_3^n)-h(Y_1^n,Y_3^n|X_r^n)\\
&= I(X_r^n;Y_1^n,Y_3^n),
\end{align*}
where $(a)$ and $(b)$ follow since conditioning does not increase entropy and $(c)$ follows from the Markov chain $\vec{m}\to X_r^n\to (Y_1^n,Y_3^n)$. The resulting mutual information expression is that of a SIMO point-to-point channel whose capacity is upper bounded by \cite{TseViswanath}
\begin{align*}
n(R_{21}+R_{31}+R_{23}-\epsilon_{n})&\leq\frac{n}{2}\log(1+(h_1^2+h_3^2)P).
\end{align*}
Thus, by letting $n\to \infty$ we get
\begin{align*}
R_{21}+R_{31}+R_{23}\leq C((h_1^2+h_3^2)P).
\end{align*}

\section{\alert{Proof of Lemma \ref{GUB}}}
\label{GeneralProof}
We prove the lemma for $(j,k,l)=(1,2,3)$, all other cases follow similarly. We shall start from Fano's inequality and give $(Y_2^n,Y_3^n,Y_r^n,m_{32})$ and $(Y_1^n,Y_2^n,Y_r^n,m_{21},m_{12},m_{13})$ as side information to receiver 1 and 3 respectively to obtain
\begin{align}
&n(R_{21}+R_{31}-\epsilon_{1n})\nonumber\\
&\quad\leq I(m_{21},m_{31};Y_1^n,Y_2^n,Y_3^n,Y_r^n,m_{32},m_{12},m_{13})\nonumber\\
\label{B2S1}
&\quad= I(m_{21},m_{31};Y_1^n,Y_2^n,Y_3^n,Y_r^n|m_{12},m_{13},m_{32}),
\end{align}
and
\begin{align}
&n(R_{23}-\epsilon_{2n})\nonumber\\
&\leq I(m_{23};Y_3^n,Y_1^n,Y_2^n,Y_r^n,m_{21},m_{12},m_{13},m_{31},m_{32})\nonumber\\
\label{B2S2}
&= I(m_{23};Y_1^n,Y_2^n,Y_3^n,Y_r^n|m_{21},m_{12},m_{13},m_{31},m_{32}).
\end{align}
Adding \eqref{B2S1} and \eqref{B2S2}, we obtain
\begin{align*}
&n(R_{21}+R_{31}+R_{23}-\epsilon_{n})\nonumber\\
&\quad\leq I(m_{21},m_{31},m_{23};Y_1^n,Y_2^n,Y_3^n,Y_r^n|m_{12},m_{13},m_{32}).
\end{align*}
Now we use the chain rule to obtain \eqref{B2CR} shown at the top of the next page, where $(a)$ follows since given $Y_r^{i-1}$, we have the necessary ingredients to construct $X_r^i$ using \eqref{RelayEncoder}.
\begin{figure*}[t]
\begin{align}
n(R_{21}+R_{31}+R_{23}-\epsilon_{n})
&\leq\sum_{i=1}^nI(m_{21},m_{31},m_{23};Y_{1i},Y_{2i},Y_{3i},Y_{ri}|Y_1^{i-1},Y_2^{i-1},Y_3^{i-1},Y_r^{i-1},m_{12},m_{13},m_{32})\nonumber\\
&\stackrel{(a)}{=}\sum_{i=1}^nI(m_{21},m_{31},m_{23};Z_{1i},Z_{2i},Z_{3i},Y_{ri}|Z_1^{i-1},Z_2^{i-1},Z_3^{i-1},Y_r^{i-1},m_{12},m_{13},m_{32})\nonumber\\
\label{B2CR}
&=I(m_{21},m_{31},m_{23};Z_{1}^n,Z_{2}^n,Z_{3}^n,Y_{r}^n|m_{12},m_{13},m_{32})
\end{align}
\hrule
\end{figure*}
Therefore, we can write
\begin{align*}
&n(R_{21}+R_{31}+R_{23}-\epsilon_{n})\nonumber\\
&\quad\leq I(m_{21},m_{31},m_{23};Z_{1}^n,Z_{2}^n,Z_{3}^n,Y_{r}^n|m_{12},m_{13},m_{32})\nonumber\\
&\quad= I(m_{21},m_{31},m_{23};Y_{r}^n|m_{12},m_{13},m_{32},Z_{1}^n,Z_{2}^n,Z_{3}^n)\nonumber
\end{align*}
by using the chain rule and the independence of the messages and the noises. Thus
\begin{align*}
&n(R_{21}+R_{31}+R_{23}-\epsilon_{n})\\
&\quad\leq \sum_{i=1}^n h(Y_{ri}|m_{12},m_{13},m_{32},Z_{1}^n,Z_{2}^n,Z_{3}^n,Y_r^{i-1})\\
&\qquad-\sum_{i=1}^nh(Y_{ri}|\vec{m},Z_{1}^n,Z_{2}^n,Z_{3}^n,Y_r^{i-1})\\
&\quad\leq \sum_{i=1}^n h(Y_{ri}|m_{12},m_{13},Z_{1}^n,Y_r^{i-1})\\
&\qquad-\sum_{i=1}^nh(Y_{ri}|\vec{m},Z_{1}^n,Z_{2}^n,Z_{3}^n,Y_r^{i-1})
\end{align*}
by the chain rule and since conditioning does not increase entropy. Now notice that given $(m_{12},m_{13},Z_{1}^n,Y_r^{i-1})$, we have all the necessary ingredients to construct $X_1^i$. Also, given $(\vec{m},Z_{1}^n,Z_{2}^n,Z_{3}^n,Y_r^{i-1})$, we have all the ingredients to construct $(X_1^i,X_2^i,X_3^i)$. Thus, we can write
\begin{align*}
&n(R_{21}+R_{31}+R_{23}-\epsilon_{n})\\
&\quad\leq \sum_{i=1}^n h(h_2X_{2i}+h_3X_{3i}+Z_{ri})\\
&\qquad-\sum_{i=1}^nh(Z_{ri}|\vec{m},Z_{1}^n,Z_{2}^n,Z_{3}^n,Z_r^{i-1})\\
&\quad= \sum_{i=1}^n [h(h_2X_{2i}+h_3X_{3i}+Z_{ri})-h(Z_{ri})].
\end{align*}
Since Gaussian distributions are differential entropy maximizers, we let $(X_{2i},X_{3i})$ be a Gaussian vector with zero mean and covariance matrix\footnote{The transmit signals of the users in the non-restricted Y-channel can be dependent.} 
\begin{equation}
\label{CovMatrix}
\Sigma(X_{2i},X_{3i})=\left(\begin{array}{cc}
P_{2i} &\rho_{23}\sqrt{P_{2i}P_{3i}}\\
\rho_{23}\sqrt{P_{2i}P_{3i}} & P_{3i}
\end{array}\right), 
\end{equation}
with $\rho_{23}\in[-1,1]$. Since $(X_{2i},X_{3i})$ have zero mean, then the variance of $h_2X_{2i}+h_3X_{3i}$ is equal to its second moment given by $\mathbb{E}[(h_2X_{2i}+h_3X_{3i})^2]=h_2^2P_{2i}+h_3^2P_{3i}+2h_2h_3\rho_{23}\sqrt{P_{2i}P_{3i}}$. Therefore
\begin{align*}
&n(R_{21}+R_{31}+R_{23}-\epsilon_{n})\\
&\quad\leq \sum_{i=1}^n \frac{1}{2}\log\left(1+h_2^2P_{2i}+h_3^2P_{3i}+2h_2h_3\rho_{23}\sqrt{P_{2i}P_{3i}}\right)\\
&\quad\stackrel{(b)}{\leq} \sum_{i=1}^n \frac{1}{2}\log\left(1+\left(\sqrt{h_2^2P_{2i}}+\sqrt{h_3^2P_{3i}}\right)^2\right)\\
&\quad\stackrel{(c)}{\leq} \frac{n}{2} \log\left(1+(|h_2|+|h_3|)^2P\right),
\end{align*}
where $(b)$ follows by using $h_2h_3\rho_{23}\leq|h_2||h_3|$ since $\rho_{23}\in[-1,1]$, and $(c)$ follows by using Jensen's inequality on a function that can be proved to be concave\footnote{Since the function $f(x)=\log(1+x)$ is concave and non-decreasing, $f((\sqrt{x}+\sqrt{y})^2)$ is concave if the function $g(x)=(\sqrt{x}+\sqrt{y})^2$ is concave as well (cf. composition rules in \cite{Boyd}). Thus it is sufficient to show that $(\sqrt{x}+\sqrt{y})^2$ is concave which can be shown to be true by checking its Hessian for example.}. Letting $n\to\infty$ we obtain the desired result
\begin{align}
\label{BG1}
R_{21}+R_{31}+R_{23}\leq C((|h_2|+|h_3|)^2P).
\end{align}

\section{\alert{The Functional Decode-and-Forward Scheme}}
\label{FDF}
In this scheme, each user transmits only in a fraction of the total duration of the transmission block, which allows each user to use a power larger than $P$ in the slots where it is active without violating the average power constraint. Namely, user 1, user 2, and user 3 can use powers
\begin{align}
\label{AvailablePower}
\frac{P}{\alpha_{12}+\alpha_{31}},\quad\frac{P}{\alpha_{12}+\alpha_{23}},\ \text{and}\ \frac{P}{\alpha_{23}+\alpha_{31}},
\end{align}
respectively. In what follows, we illustrate the scheme for slots $3b+1$, $3b+2$, and $3b+3$. We remove the block index from the messages for readability.

\subsection{Codebook Generation}
The users use nested-lattice codes for the communication. Here, we introduce the necessary notation required for describing the scheme, more details on nested-lattice codes can be found in \cite{NazerGastpar}. A nested-lattice code is constructed from a fine lattice $\Lambda^f$ and a coarse lattice $\Lambda^c\subset\Lambda^f$, and is denoted by the pair $(\Lambda^f,\Lambda^c)$. For $\lambda_1,\lambda_2\in\Lambda_f$, we have
\begin{align*}
\lambda_1+\lambda_2\in\Lambda^f.
\end{align*}
A nested-lattice codebook consists of the points $\lambda\in\Lambda^f\cap\mathcal{V}(\Lambda^c)$ where $\mathcal{V}(\Lambda^c)$ is the fundamental Voronoi region of $\Lambda^c$, defined as the set of $\vec{x}\in\mathbb{R}^n$ such that the distance between $\vec{x}$ and the all zero vector is smaller than that to any other point in  $\Lambda^c$. The power constraint is satisfied by an appropriate choice of $\Lambda^c$ and the rate of the code is defined by the number of fine lattice points in $\Lambda^f\cap\mathcal{V}(\Lambda^c)$. 

In the sequel, we need the following lemma. Consider $\lambda_A$ and $\lambda_B$, two length-$n$-signals which are points in the nested-lattice codebook $(\Lambda_f,\Lambda_c)$ with rate $R$ and power $P$. Let
\begin{align}
\label{XA}
x_A^n&=(\lambda_A-d_A)\mod\Lambda_c\\
\label{XB}
x_B^n&=(\lambda_B-d_B)\mod\Lambda_c
\end{align}
where $d_A$ and $d_B$ are $n$-dimensional random dithers\cite{NazerGastpar}. 
\begin{lemma}
\label{Lemma:FDF}
Assume that a node receives the signal $y^n=x_A^n+x_B^n+z^n$ where $z^n$ is a noise sequence with i.i.d. components $\mathcal{N}(0,\sigma^2)$. If this node knows the random dithers, then it can decode the sum $(\lambda_A+\lambda_B)\mod \Lambda_c$ from $y^n$ reliably as long as \cite{NarayananPravinSprintson} $$R\leq\frac{1}{2}\log\left(\frac{1}{2}+\frac{P}{\sigma^2}\right).$$ 

Additionally, if a node knows $(\lambda_A+\lambda_B)\mod \Lambda_c$ and $\lambda_A$, it can extract $\lambda_B$\cite{NarayananPravinSprintson}.
\end{lemma}

Now we can proceed with describing our scheme. Consider time slot 1. The nested-lattice code used for encoding messages $m_{12}$ and $m_{21}$ are $(\Lambda_{12}^f,\Lambda_{12}^c)$ and $(\Lambda_{21}^f,\Lambda_{21}^c)$, respectively. Both nested-lattice codes have rate $R_{12}$ for simplicity of exposition. The nested-lattice code $(\Lambda_{21}^f,\Lambda_{21}^c)$ satisfies
\begin{align}
\label{Align1}
(h_1\Lambda^f_{12},h_1\Lambda^c_{12})&=\left(h_2\Lambda^f_{21},h_2\Lambda^c_{21}\right),
\end{align}
and hence, it has power 
\begin{align}
\label{PowerAlignment}
P_{21}&=\frac{h_1^2}{h_2^2}P_{12}.
\end{align}

Notice that this guarantees that the nested-lattice codes are aligned at the relay\footnote{\alert{The nested-lattice scheme introduced in \cite{NamChungLee_IT} is able to handle different powers/channel gains in the uplink. This scheme can also be used for approximating the sum-capacity of the Y-channel.}}, i.e., the relay receives the sum of two signals both of which correspond to the nested-lattice code $(h_1\Lambda^f_{12},h_1\Lambda^c_{12})$. This allows the relay to decode the superposition of codewords (see Lemma \ref{Lemma:FDF}). Similar construction is used for slots 2 and 3, where the rates are $R_{23}$ and $R_{31}$, and the powers $P_{32}=\frac{h_2^2}{h_3^2}P_{23}$ and $P_{31}=\frac{h_1^2}{h_3^2}P_{31}$, respectively.

The relay uses three Gaussian codebooks of rates $R_{12}$, $R_{23}$, and $R_{31}$ and power $P$. For instance, for the communication between nodes 1 and 2, the relay sends a message $u_{12}\in\mathcal{U}_{12}\triangleq\{1,\dots,2^{nR_{12}}\}$ where the subscripts indicate that the message carries information to both users $1$ and $2$. \alert{This messages $u_{12}$ is an index which corresponds to the decodes sum of codewords at the relay as we shall see next.}

\subsection{Encoding at the Sources}
Consider slot $3b+1$. Users 1 and 2 send $\alpha_{12}n$ symbols of
$$x_1^n=(\lambda_{12}-d_{12})\mod \Lambda^c_{12},$$ and $$x_2^n=(\lambda_{21}-d_{21})\mod \Lambda^c_{21},$$ respectively, using their nested-lattice codes as in \eqref{XA} and \eqref{XB}. User 3 is kept silent in this slot. A similar encoding is used in slots 2 and 3.

\subsection{Processing at the Relay}
The relay collects $n$ symbols (code length) corresponding to communication between users 1 and 2. The resulting collection can be written as 
\begin{align*}
y_r^n&=h_1x_1^n+h_2x_2^n+z_r^n.
\end{align*}
The relay can then decode the modulo-coarse-lattice sum $(h_1\lambda_{12}+h_2\lambda_{21})\mod h_2\Lambda_{21}^c$, with arbitrarily small probability of error (see Lemma \ref{Lemma:FDF}) if
\begin{align*}
R_{12}=R_{21}&\leq C^+\left(h_2^2P_{21}-\frac{1}{2}\right).
\end{align*}
In order to maximize this expression, we should choose the largest $P_{21}$, denoted $P_{21}^*$, such that the power constraint is satisfied at user 1 and 2, i.e., 
\begin{align*}
\frac{h_2^2}{h_1^2}P_{21}^*&\leq \frac{P}{\alpha_{12}+\alpha_{31}},\\
P_{21}^*&\leq \frac{P}{\alpha_{12}+\alpha_{23}},
\end{align*}
where we used \eqref{AvailablePower} and \eqref{PowerAlignment}. Therefore, we get \eqref{P21*} leading to the following rates
\begin{align*}
R_{12}=R_{21}\leq C^+\left(h_2^2P_{21}^*-\frac{1}{2}\right).
\end{align*}
By similar analysis, we get \eqref{P31*} and \eqref{P32*}. Notice that this choice of powers satisfies the power constraints at all users, keeping in mind that each user is active for 2 slots out of 3 slots.

After decoding $(h_1\lambda_{12}+h_2\lambda_{21})\mod h_2\Lambda_{21}^c$, the relay maps it to an index $u_{12}\in\mathcal{U}_{12}$. Then, it maps $u_{12}$ into a codeword $x_r^n(u_{12})$, and transmits $x_r^n(u_{12})$. Keep in mind that this message $u_{12}$ is meant for users 1 and 2.

\subsection{Decoding at the Destinations}
User 1 has the received signal $y_1^n=h_1x_r^n+z_1^n$. The relay index $u_{12}$ is then decoded, and $m_{21}$ is extracted (see Lemma \ref{Lemma:FDF}). This can be done with an arbitrarily small probability of error if 
\begin{align*}
R_{12}=R_{21}&\leq C\left(h_1^2P\right).
\end{align*}
User 2 obtains $m_{12}$ reliably if 
\begin{align*}
R_{12}=R_{21}&\leq C\left(h_2^2P\right).
\end{align*}
Similarly, users 2 and 3 decode $m_{32}$ and $m_{23}$, and users 1 and 3 decode $m_{31}$ and $m_{13}$.

\subsection{Rate Constraints}
As a result, the achievable rates using this scheme are bounded by
\begin{align*}
R_{12}=R_{21}&\leq\min\left\{C^+\left(h_2^2P_{21}^*-\frac{1}{2}\right),C\left(h_2^2P\right)\right\}\\
R_{13}=R_{31}&\leq\min\left\{C^+\left(h_3^2P_{31}^*-\frac{1}{2}\right),C\left(h_3^2P\right)\right\}\\
R_{23}=R_{32}&\leq\min\left\{C^+\left(h_3^2P_{32}^*-\frac{1}{2}\right),C\left(h_3^2P\right)\right\}.
\end{align*}
Since the slots have length $\alpha_{12}n$, $\alpha_{23}n$, and $\alpha_{31}n$, we obtain the following achievable sum-rate

\begin{align*}
R_\Sigma&=2\alpha_{12}\min\left\{C^+\left(h_2^2P_{21}^*-\frac{1}{2}\right),C\left(h_2^2P\right)\right\}\\
&+2\alpha_{23}\min\left\{C^+\left(h_3^2P_{31}^*-\frac{1}{2}\right),C\left(h_3^2P\right)\right\}\\
&+2\alpha_{31}\min\left\{C^+\left(h_3^2P_{32}^*-\frac{1}{2}\right),C\left(h_3^2P\right)\right\}.
\end{align*}

\section{Proof of Lemma \ref{RUB}}
\label{ProofRestricted}
All the steps of the proof of Lemma \ref{GUB} for the non-restricted Y-channel go through for the restricted Y-channel in a similar way. However, for the restricted case, the random variables $X_{2i}$ and $X_{3i}$ are independent, leading to 
\begin{equation*}
\Sigma(X_{2i},X_{3i})=
\left(\begin{array}{cc}
P_{2i} &0\\
0 & P_{3i}
\end{array}\right), 
\end{equation*}
instead of \eqref{CovMatrix}. Thus, we get
\begin{align*}
&n(R_{21}+R_{31}+R_{23}-\epsilon_n)\\
&\quad\leq \sum_{i=1}^n h(h_2X_{2i}+h_3X_{3i}+Z_{ri})-h(Z_{ri})\\
&\quad\leq \sum_{i=1}^n \frac{1}{2}\log\left(1+h_2^2P_{2i}+h_3^2P_{3i}\right)\\
&\quad\leq \frac{n}{2} \log\left(1+(h_2^2+h_3^2)P\right),
\end{align*}
which follows by using Jensen's inequality. Then, we can bound $R_{21}+R_{31}+R_{23}$ by
\begin{align*}
R_{21}+R_{31}+R_{23}\leq C((h_2^2+h_3^2)P).
\end{align*}

\section{Upper Bounds for the K-user Star Channel}
\label{Proof:KUserStar}
Let us start be defining some notation. Denote by $\mathcal{M}_j$ the set of messages originating at user $j$, i.e., $$\mathcal{M}_j=\{m_{j,1},\dots,m_{j,j-1},m_{j,j+1}\dots,m_{j,K}\}.$$ Similarly, denote by $\widehat{\mathcal{M}}_j$ the set of messages intended to user $j$, i.e., 
$$\widehat{\mathcal{M}}_j=\{m_{1,j},\dots,m_{j-1,j},m_{j+1,j}\dots,m_{K,j}\}.$$

Consider user $K$. The rate of information flow to user $K$ can be bounded by\footnote{$R_{j,K}$ is that same as $R_{jK}$, we introduced a comma in the subscripts for readability.}
\begin{align}
\label{BoundK}
n\sum_{j=1}^{K-1}R_{j,K}\leq I(\widehat{\mathcal{M}}_K;Y_K^n,\mathcal{M}_K)+n\epsilon_n
\end{align}
by using Fano's inequality, where $\epsilon_n\to0$ as $n\to\infty$. From this point on, we will omit $\epsilon_n$ for ease of exposition. This upper bound can be used to obtain 
\begin{align}
\label{BoundKH}
\sum_{j=1}^{K-1}R_{j,K}\leq\frac{1}{2}\log(P)+o(\log(P)).
\end{align}
Now we need to find a bound which can be added to \eqref{BoundK} without changing the asymptotic behavior of the bound (i.e., \eqref{BoundKH}). Consider user $K-1$, and assume that user $K$ passes all its information after decoding to user $K-1$. Then, for this user we can write
\begin{align}
\label{BoundK-1}
n\hspace{-0.12cm}\sum_{j=1}^{K-2}\hspace{-0.12cm}R_{j,K-1}\hspace{-0.1cm}\leq I(\widehat{\mathcal{M}}_{K-1\setminus K};Y_K^n,Y_{K-1}^n,\mathcal{M}_{K},\mathcal{M}_{K-1\setminus K}|\widehat{\mathcal{M}}_K)
\end{align}
where $\mathcal{M}_{K-1\setminus K}$ and $\widehat{\mathcal{M}}_{K-1\setminus K}$ are used as shorthand notations for $\mathcal{M}_{K-1}\setminus\{m_{K-1,K}\}$ and $\widehat{\mathcal{M}}_{K-1}\setminus\{m_{K,K-1}\}$, respectively, i.e., the set of messages originating at user $K-1$ except the one intended to user $K$, and the set of messages intended to user $K-1$ except the one originating at user $K$. In order to be able to add \eqref{BoundK} and \eqref{BoundK-1}, we give $(Y_{K-1}^n,\mathcal{M}_{K-1\setminus K})$ to user $K$ to obtain
\begin{align}
\label{BoundK_2}
n\sum_{j=1}^{K-1}R_{j,K}\leq I(\widehat{\mathcal{M}}_K;Y_K^n,Y_{K-1}^n,\mathcal{M}_K,\mathcal{M}_{K-1\setminus K}).
\end{align}
Now, \eqref{BoundK-1} and \eqref{BoundK_2} can be added by using the chain rule of mutual information to obtain \eqref{BoundK+K-1} at the top of next page. 
\begin{figure*}[t]
\begin{align}
\label{BoundK+K-1}
n\sum_{k=K-1}^{K}\sum_{j=1}^{k-1}R_{j,k}&\leq I\left(\bigcup_{\ell=K-1}^K\widehat{\mathcal{M}}_{\ell\setminus \ell+1};Y_K^n,Y_{K-1}^n,\bigcup_{\ell=K-1}^K\mathcal{M}_{\ell\setminus\ell+1}\right)\\
\label{BoundAll}
n\sum_{k=2}^{K}\sum_{j=1}^{k-1}R_{j,k}&\leq I\left(\bigcup_{\ell=2}^K\widehat{\mathcal{M}}_{\ell\setminus [\ell+1:K]};Y_K^n,\dots,Y_{2}^n,\bigcup_{\ell=2}^K\mathcal{M}_{\ell\setminus[\ell+1:K]}\right)\\
\label{BoundAll2}
n\sum_{k=1}^{K-1}\sum_{j=k+1}^{K}R_{j,k}&\leq I\left(\bigcup_{\ell=1}^{K-1}\widehat{\mathcal{M}}_{\ell\setminus [1:\ell-1]};Y_1^n,\dots,Y_{K-1}^n,Y_r^n,\bigcup_{\ell=1}^{K-1}\mathcal{M}_{\ell\setminus[1:\ell-1]}\right)
\end{align}
\hrule
\end{figure*}
Notice that since $Y_K^n$ is a degraded version of $Y_{K-1}^n$ assuming that user $K-1$ has a better channel (without loss of generality), then \eqref{BoundK+K-1} still has the same behavior as \eqref{BoundKH}. By repeating this process similarly for users $K-2\dots2$, every time giving the necessary side information to the users, we can obtain \eqref{BoundAll} at the top of next page, where $\widehat{\mathcal{M}}_{\ell\setminus [\ell+1:K]}$ denotes the set of messages intended to user $\ell$ except those originating at users $\ell+1,\dots,K$, such that $\widehat{\mathcal{M}}_{K\setminus [K+1:K]}=\widehat{\mathcal{M}}_{K}$, and $\mathcal{M}_{\ell\setminus[\ell+1:K]}$ is defined similarly. Then we can proceed as follows
\begin{align*}
n\sum_{k=2}^{K}\sum_{j=1}^{k-1}R_{j,k}&\leq h\left(Y_K^n,\dots,Y_{2}^n|\bigcup_{\ell=2}^K\mathcal{M}_{\ell\setminus[\ell+1:K]}\right)\\
&\quad-h\left(Y_K^n,\dots,Y_{2}^n|\bigcup_{\ell=1}^K\mathcal{M}_{\ell}\right)\\
&\leq h\left(Y_K^n,\dots,Y_{2}^n\right)-h\left(Y_K^n,\dots,Y_{2}^n|\bigcup_{\ell=1}^K\mathcal{M}_{\ell}\right)\nonumber\\
&\leq h\left(Y_K^n,\dots,Y_{2}^n\right)-h\left(Y_K^n,\dots,Y_{2}^n|X_r^n\right)\nonumber\\
&= I\left(X_r^n;Y_K^n,\dots,Y_{2}^n\right).
\end{align*}
The resulting expression resembles a SIMO point-to-point channel, whose upper bound can be written as \cite{TseViswanath}
\begin{align*}
\sum_{k=2}^{K}\sum_{j=1}^{k-1}R_{j,k}&\leq \frac{1}{2}\log(1+(\|\vec{h}\|_2)^2P),
\end{align*}
where $\vec{h}$ is the vector $(h_2,h_3,\dots,h_K)$ and $\|\vec{h}\|_2$ is its $\ell^2$-norm. Similarly, we can obtain \eqref{BoundAll2} using similar steps as those in Appendix \ref{GeneralProof}, where $\widehat{\mathcal{M}}_{1\setminus [1:0]}=\widehat{\mathcal{M}}_{1}$. Namely, we start with
\begin{align*}
n\sum_{j=2}^{K}R_{j,1}&\leq I\left(\widehat{\mathcal{M}}_{1};Y_1^n,Y_r^n,\mathcal{M}_{1}\right)\\
n\sum_{j=3}^{K}R_{j,2}&\leq I\left(\widehat{\mathcal{M}}_{2\setminus 1};Y_2^n,Y_r^n,\mathcal{M}_{2}\right).
\end{align*}
We notice that if we give $Y_2^n$ and $\mathcal{M}_{2\setminus 1}$ to user 1 and $Y_1^n$, $\mathcal{M}_{1}$, and $\widehat{\mathcal{M}}_{1}$ to user 2 as side information, then we can add the two bounds using the chain rule of mutual information to obtain
\begin{align*}
&n\sum_{k=1}^{2}\sum_{j=k+1}^{K}R_{j,k}\\
&\quad\leq I\left(\bigcup_{\ell=1}^{2}\widehat{\mathcal{M}}_{\ell\setminus [1:\ell-1]};Y_1^n,Y_{2}^n,Y_r^n,\bigcup_{\ell=1}^{2}\mathcal{M}_{\ell\setminus[1:\ell-1]}\right).
\end{align*}
By proceeding with similar steps, considering one more user each time, we end up with \eqref{BoundAll2}, from which we can obtain
\begin{align*}
\sum_{k=1}^{K-1}\sum_{j=k+1}^{K}R_{j,k}&\leq \frac{1}{2}\log\left(1+\left(\sum_{\ell=2}^K|h_\ell|\right)^2P\right)\\
&= \frac{1}{2}\log\left(1+(\|\vec{h}\|_1)^2P\right),
\end{align*}
using similar steps as those used from \eqref{B2CR} to \eqref{BG1} in the 3-user case in Appendix \ref{GeneralProof}, where $\|\vec{h}\|_1$ is the $\ell^1$-norm of $\vec{h}$.

\section{Gap Calculation for the $K$-user Case}
\label{Proof:KUserStar_Gap}
We start by repeating the upper bound \eqref{KUserUB} which can be rewritten as
\begin{align*}
C_\Sigma&\leq \frac{1}{2}\log\left(1+\sum_{\ell=2}^Kh_\ell^2P\right)+ \frac{1}{2}\log\left(1+\left(\sum_{\ell=2}^K|h_\ell|\right)^2P\right).
\end{align*}
From this bound, we can obtain
\begin{align*}
C_\Sigma&\leq \frac{1}{2}\log\left(1+(K-1)h_2^2P\right)+ \frac{1}{2}\log\left(1+(K-1)^2h_2^2P\right)\nonumber\\
&\leq \log\left(1+(K-1)^2h_2^2P\right)\\
&\leq \log\left(1+2h_2^2P\right)+2\log\left(K-1\right)-1.
\end{align*}

For $h_2^2P>\frac{1}{2}$, the lower bound $\underline{C}_3$ from Corollary \ref{LowerBound:FDF_2User} is given by
\begin{align*}
\underline{C}_{3}&\geq2C\left(h_2^2P-\frac{1}{2}\right)\\
&=\log\left(1+2h_2^2P\right)-1.
\end{align*}

Now, by comparing the obtained upper and lower bounds, we can deduce that the gap between them is $2\log\left(K-1\right)$, and thus, we have the sum-capacity of the network within $\mathcal{O}(\log(K))$.

\end{appendices}

\section*{Acknowledgement}
The authors would like to express their gratitude to the reviewers and the editor for helpful comments which improved the quality of the paper.


\end{document}